\documentclass[12pt,notitlepage]{revtex4-1}
\usepackage{graphicx}
\usepackage{amsmath}
\usepackage{amsfonts}
\usepackage{amssymb}
\usepackage{epstopdf}

\font\nota=cmr8

\let\a=\alpha   \let\g=\gamma  \let\d=\delta \let\e=\varepsilon
\let\z=\zeta  \let\h=\eta    \let\k=\kappa 
        \let\x=\xi         \let\r=\rho
\let\s=\sigma    \let\f=\varphi \let\c=\chi
   
\let\G=\Gamma \let\D=\Delta  \let\L=\Lambda 
         
\let\O=\Omega \let\Y=\Upsilon

\def\PPP{{\cal P}} 
\def\CC{{\cal C}}\def\FF{{\cal F}} 
\def\NN{{\cal N}} 
  
\def\GG{{\cal G}} \def\SS{{\cal S}}

\def\Ft{\varphi}

 \def\ZZZ{{\mathbb Z}}

\def\\{\hfill\break} 
\let\==\equiv
 
\let\0=\noindent
\def\media#1{{\langle#1\rangle}}
\let\dpr=\partial

\def\const{{\rm const}}
\def\tende#1{\,\vtop{\ialign{##\crcr\rightarrowfill\crcr\noalign{\kern-1pt
    \nointerlineskip} \hskip3.pt${\scriptstyle #1}$\hskip3.pt\crcr}}\,}
\def\otto{\,{\kern-1.truept\leftarrow\kern-5.truept\to\kern-1.truept}\,}

\def\to{\rightarrow}

\def\qed{\hfill\raise1pt\hbox{\vrule height5pt width5pt depth0pt}}

\def\ul#1{{\underline#1}}
\def\lis{\overline}

\def\be{\begin{equation}}
\def\ee{\end{equation}}
\def\bea{\begin{eqnarray}}
\def\eea{\end{eqnarray}}
\def\nn{\nonumber}

\def\fra#1#2{\frac{#1}{#2}}

\def\supp{{\rm supp}}
\def\dist{{\rm dist}}
\def\Ext{{\rm Ext}}
\def\Int{{\rm Int}}
\def\diam{{\rm diam}}

\newtheorem{theor}{\large\bf Theorem}
\newtheorem{lemma}{Lemma}

\begin{document}

\title{The nematic phase of a system of long hard rods}
\thanks{\copyright\, 2011  by the authors. This paper may be
reproduced, in its entirety, for non-commercial purposes.}
\author{Margherita Disertori}
\affiliation{Laboratoire de Math\'ematiques Rapha\"el Salem, 
UMR 6085 CNRS-Universit\'e de Rouen, Av. de l'Universit\'e, BP.12,
Technop\^ole du Madrillet, 
F76801 Saint-\'Etienne-du-Rouvray -  France}
\author{Alessandro Giuliani}
\affiliation{Dipartimento di Matematica, Universit\`a di Roma Tre,
L.go S. L. Murialdo 1, 00146 Roma - Italy}
\vspace{3.truecm}

\begin{abstract} We consider a two-dimensional lattice model for liquid crystals consisting of long rods interacting via purely hard core interactions, with two allowed orientations defined by the underlying lattice. 
We rigorously prove the existence of a nematic phase, i.e., we show that at intermediate 
densities the system exhibits orientational order, either horizontal or vertical, but no positional order. 
The proof is based on a two-scales cluster expansion: we first coarse grain the 
system on a scale comparable with the rods' length; then we express the resulting effective theory as a 
contour's model, which can be treated by Pirogov-Sinai methods. 
\end{abstract}

\maketitle

\renewcommand{\thesection}{\arabic{section}}

\vskip.2truecm

{\it Dedicated to the 70th birthday of Giovanni Gallavotti}

\section{Introduction}\label{sec1}
\setcounter{equation}{0}
\renewcommand{\theequation}{\ref{sec1}.\arabic{equation}} 


In 1949, L. Onsager \cite{O} proposed a statistical theory for a system of elongated molecules
interacting via repulsive short-range forces, based on an explicit computation of the first few Mayer's
coefficients for the pressure. Onsager's theory predicted the existence  at intermediate densities of 
a nematic liquid crystal phase, that is a phase in which the distribution of orientations of the particles 
is anisotropic, while the distribution of the particles in space is homogeneous and does not exhibit the 
periodic variation of densities that characterizes solid crystals (periodicity in all space dimensions)
or smectic liquid crystals (periodicity in one dimension).

From a microscopic point of view, the most natural lattice model describing elongated molecules 
with short-range repulsive forces is a system of rods of length $k$ and thickness 1 at fixed density $\r$ 
(here $\r=$ average number of rods per unit volume), 
arranged on a cubic lattice, say a large squared box portion of 
$\ZZZ^2$, and interacting via a purely hard core potential. Even though very natural,
this model is not easy to treat and its  
phase diagram in the plane $(\r,k)$ is still not understood in many physically relevant 
parameters' ranges. Of course, for all $k$'s, at very small density there is a unique isotropic
Gibbs state,  invariant under translations and under discrete rotations of $90^o$; this can be proved by standard cluster expansion methods. If $k=2$, it is known 
\cite{HL1} that the state is analytic and, therefore, there is no phase transition, for all densities 
but, possibly, at the close packing density, i.e., at the maximal possible density 
$\r_{max}=1/k$. If $k$ is sufficiently large ($k\ge 7$ should be enough \cite{GD})
there is numerical evidence \cite{GD,MLR} for {\it two} phase transitions as $\r$ is increased from 
zero to the maximal density. The first, isotropic to nematic, seems to take place 
at a $\r_c^{(1)}\simeq C_1/k^2$, while the second, nematic to isotropic, seems to take place
at $\r_c^{(2)}\simeq \r_{max}-C_2/k^3$. These findings renovated the interest of the condensed matter community in the phase diagram of long hard rod systems and stimulated more systematic numerical 
studies of the nature of the critical points at $\r_c^{(1)}$ and $\r_c^{(2)}$
\cite{DRS,FV,LLR,LLRC,MLR2,MLR3}.
From a mathematical point of view
there is no rigorous proof of any of these behaviors yet, with the exception of the ``trivial" case of very low densities: namely, there is neither a proof of nematic 
order at intermediate densities, nor a proof of the absence of orientational order at very high densities,
nor a rigorous understanding of the nature of the transitions. 

In this work we give a rigorous proof to some of the conjectures stated above on the nature of the 
phase diagram of long hard rods systems. More precisely, we show that well inside the interval 
$(\r_c^{(1)},\r_c^{(2)})$, the system is in a nematic phase, i.e., in a phase characterized by two 
distinguished Gibbs states, with different orientational order, either horizontal or vertical, but with no 
positional order. 
To the best of our knowledge, this is the first proof of the existence of a nematic 
phase in a microscopic model with molecules of fixed finite length and finite thickness, 
interacting via a purely repulsive potential. In this respect, our result is a strong confirmation of 
Onsager's proposal that orientational ordering can be explained as an excluded volume effect.

Our proof is based on a two-scales cluster expansion 
method, in which we first coarse grain the system on scale $k$; we next realize that the resulting 
effective model can be expressed as a contour model, reminiscent of the contour theory 
for the Ising model at low temperatures. However, contrary to the Ising case, the contour theory we have to deal with here is not invariant 
under a ${\mathbb Z}^2$ symmetry: therefore, we cannot apply the Peierls' argument and we need to 
make use of a Pirogov-Sinai method.

Of course, our proof leaves many questions about the phase diagram of long hard rod systems open,
the most urgent being, we believe, the question about the nature of the densely packed phase at $\r\ge 
\r_{c}^{(2)}$: can one prove the absence of orientational order, at least at close packing? 
Is the densely packed phase characterized by some ``hidden" (striped-like) order? Progress
on these problems would be important for the understanding of the emergence of hidden order in 
more complicated systems than elongated molecules with purely hard core interactions, in which short 
range repulsion competes with attractive forces acting on much longer length scales. \\

{\it Previous results.} 
There is a limited number of papers where important previous results on the existence of orientational order in lattice or continuum models for liquid crystals were obtained, related to the ones found in this work. 

A first class of liquid crystal models that has been considered in the literature describe long rods with purely repulsive interactions and discrete orientations, like ours; of course, the case of continuous orientations would be of great interest, but its treatment appears to be beyond the current state of the art. 
In \cite{He,Hu,LG}, the existence of orientational order for different variants of lattice gases of anisotropic molecules with 
repulsive interactions was proved, by using Peierls-like estimates and cluster expansion; however, in all these cases, orientational comes together with translational order, which is not the case in a nematic phase. A continuum version of the model 
in \cite{LG}, i.e., a continuum system of infinitely thin 
rods with two allowed orientations and hard core interactions, was later proved to have a phase transition from an isotropic to a nematic phase \cite{Ru71,BKL}, by using improved estimates on the 
contours' probabilities and a Pirogov-Sinai method. 
More recently, the existence of an isotropic to nematic transition in an 
integrable model of polydisperse long rods in $\ZZZ^2$ with hard core interactions was proved \cite{IVZ}, by mapping the partition function of the polydisperse hard rods gas into that of the 
nearest neighbor 2D Ising model. 

A second class of liquid crystal models studied in the literature assumes the existence of 
attractive forces favoring the alignment of the molecular axes: in fact, in some cases, the attraction is 
expected to originate from the inter-molecular Coulomb interaction \cite{MS} and to play a more 
prominent role than the Onsager's excluded volume effect. The emergence of a nematic phase 
in such models was first understood at the mean field level \cite{BZ, dGP,MS, P}. 
Later, it was understood that in the presence of attractive forces, 
even the monomer-dimer system can exhibit an oriented phase at low temperatures, as proved
in \cite{HL2} by reflection positivity methods; the absence of positional order for the same model,
known as the Heilmann-Lieb's model \cite{HL2}, was then proved on the basis of cluster expansion 
methods \cite{L}. 
Remarkably, if attractive forces favoring the alignment of the molecules' axes are allowed,
there are models displaying a full $O(m)$ orientational symmetry, $m\ge2$, for which it is possible to 
rigorously prove the existence of nematic order (or quasi-long range order, depending on the dimensionality). In particular, in \cite{AZ, ARZ, Za} certain $d$-dimensional lattice-gas models describing particles with 
an internal (``spin") continuous orientational degree of freedom were introduced; 
the existence of orientational order was proved both in $d>2$ for short-ranged interactions and 
in $d=1,2$ with sufficiently long-ranged interactions, via a combination of infrared bounds and chessboard estimates \cite{FILS}. In  \cite{GTZ}, a proof of the existence of 
orientational quasi long range order a'la Kosterlitz-Thouless 
was given for a similar system in $d=2$ with short ranged interactions, by using a combination of the Gruber-Griffiths method \cite{GG}, originally applied to 
the study of an orientational phase transition in a continuum system of
particles with internal Ising-like degrees of freedom, and of the Fr\"ohlich-Spencer method \cite{FS}, originally 
applied to the study of the Kosterlitz-Thouless transition in the classical two-dimensional XY model.
\\

{\it Summary.}
The rest of the paper is organized as follows. In Section \ref{sec2} we ``informally" introduce the model, 
state the main results and explain the key ideas involved in the proof. In Section \ref{sec3} we define
the model and state the main theorem (Theorem \ref{thm1} below) in a mathematically precise form.
In the following sections we prove Theorem \ref{thm1}: in Section \ref{sec4} we 
rewrite the partition function with $q$ boundary conditions in terms of a sum over contours' 
configurations, where the contours are defined in a way suitable for later application of a Pirogov-Sinai 
argument. In Section \ref{sec5} we prove the convergence of the cluster expansion for the 
pressure, under the assumption that the activity of the contours is small and decays sufficiently fast in the contour's size. 
In Section \ref{sec6} we complete the proof of convergence of the cluster expansion for the 
pressure, by inductively proving the desired bound on the activity of the contours.
Finally, in Section \ref{sec7} we adapt our expansion to the computation of correlation functions and
we prove Theorem \ref{thm1}.


\section{The model}\label{sec2}
\setcounter{equation}{0}
\renewcommand{\theequation}{\ref{sec2}.\arabic{equation}} 


We consider a finite square box $\L\subset \ZZZ^2$ of side $L$, to be 
eventually sent to infinity. 
We fix $k$ and the average density $\r\in(0,1/k)$.
The finite volume Gibbs measure at activity $z$ gives weight 
$z^n$ to every allowed configuration of $n$ rods: we say that a configuration is allowed if
no pair of rods overlaps. Of course, one also needs to specify boundary conditions: we consider, say,
periodic boundary conditions, open boundary conditions, horizontal or vertical boundary 
conditions, the latter meaning that all the rods within a distance $\sim k$ from the boundary of
$\L$ are horizontal or vertical -- see below for a more precise definition.
The grand canonical partition function is:
\be Z_\L(z)=\sum_{n\ge 0}z^n w^\L_n 
\;,\label{2.1}\ee
where $w^\L_n$ is the number of allowed configurations of $n$ rods in the box $\L$, in the 
presence of the prescribed boundary conditions. Note that $w^\L_n=0$ for all $n\ge |\L|/k$, which 
shows that $Z_\L(z)$ is a finite (and, therefore, well defined) sum for all finite $\L$'s. 
The activity $z$ is fixed in such a way that 
\be \lim_{|\L|\to\infty}\frac{\media{n}_\L}{|\L|}=\lim_{|\L|\to\infty}\frac{1}{|\L|} 
\frac{\sum_{n\ge 0}n z^n w^\L_n  }{Z_\L(z)}=\r\;.\label{2.2}\ee
The goal is to understand the properties of the partition function and of 
the associated Gibbs state in the limit $|\L|\to \infty$ at fixed $\r$.
An informal statement of our main result is the following.\medskip

{\bf Main result.} {\it For $k$ large enough, if $k^{-2}\ll \r \ll k^{-1}$, 
the system admits two distinct  infinite volume Gibbs states, characterized by long 
range orientational order (either horizontal or vertical) and no translational order, 
selected  by the boundary conditions.}
\medskip

{\bf Sketch of the proof.} The idea is to  coarse grain  $\Lambda $
in squares of side  $\ell\simeq k/2$. Each square is large, since in average it 
contains many ($ \sim \r k^2 \gg 1$) rods. On the other hand, 
its side $\ell $ is small enough to ensure
that only rods of the same orientation are allowed to have centers in the same square.
This means that the partition function restricted to a single square
contains only sums over vertical or horizontal configurations.
Let us consider the case where the rods  are all horizontal (vertical is treated
in the same way). A typical horizontal configuration consists of many ($\sim \r k^2$) 
horizontal rods with centers distributed 
approximately uniformly (Poisson-like) in the square, since their interaction, 
once we prescribe their  direction, is very weak: they ``just" have a hard core 
repulsion that prevents two rods to occupy the same row, an event that is very rare, 
since the density of occupied rows ($\sim \r k^2/k$)
is very small, thanks to the condition that $\r \ll1/k$. Because of this small density of 
occupied rows,  we are able to quantify via cluster expansion methods how close to 
Poissonian is the distribution of the centers in the given square (once we condition with respect to
a prescribed orientation of the rods). 

To control the interaction between different squares we use a Pirogov-Sinai argument. 
Each square can be of three types:  
(i) either it is of type $+1$, if it contains only horizontal rods, (ii) or it is of type $-1$, if 
it contains only vertical rods, (iii) or it is of type $0$, if it is empty. 
The values $-1,0,+1$ associated to each square play the role of spin values associated to the 
coarse grained system. The interaction between the 
spins is only finite range and squares with vertical ($+1$) and horizontal ($-1$) spin have a strong repulsive interaction, due to the hard core constraint. On the other hand, the vacuum configurations 
(the spins equal to 0) are very unlikely, since the probability of having a large deviation event such 
that a square of side 
$\ell$ is empty is expected to be exponentially small $\sim \exp\{-c\r k^2\}$, for a suitable constant 
$c$.

Therefore the typical spin
configurations consist of big connected clusters of ``uniformly magnetized spins", either of type 
$+1$ or of type $-1$ separated by boundary layers (the contours), which contain 
zeros or pairs of neighboring opposite spins. These contours can be shown to satisfy a Peierls' 
condition, i.e., the probability that a given contour occurs is exponentially small in the 
size of its geometric support. 
The contour theory is not symmetric under spin flip and, therefore, we are forced
to study it by the (non-trivial although standard) methods first introduced
by  Pirogov and Sinai \cite{PS}. \\

Before we move to discuss the details of our proof, let us state 
our main results in a mathematically more sound form. 


\section{Main results}\label{sec3}
\setcounter{equation}{0}
\renewcommand{\theequation}{\ref{sec3}.\arabic{equation}} 


{\bf Definitions.} For any region 
 $X \subseteq\mathbb{Z}^{2}$  we call $\Omega_{X}$ the
set of rod configurations $R=\{r_{1},\dotsc ,r_{n}\}$  where all the
rods belong to the region $X$. A rod $r$ ``{\em belongs to}" a region $X$
if the center of the rod is inside the region, in which case we write $r\in X$.
Here each rod is identified with a sequence of $k$ adjacent sites of $\ZZZ^2$ in the horizontal or 
vertical direction. If $k$ is odd, the center of the rod belongs to the
lattice $\ZZZ^2$ itself and, therefore, the notion of ``rod belonging to $X$"
is unambiguously defined. On the contrary, if $k$ is even, the geometrical center of the rod does
not belong to the original lattice $\ZZZ^2$; however, for what follows, 
it is convenient to pick one of the sites belonging to $r$ and elect it to the role of 
``center of the rod": if $r$ is horizontal (vertical), we decide that the ``center of $r$" 
is the site of $r$ that is closest to its geometrical center from the left 
(bottom). We shall also say that: a rod $r$ 
``{\em touches}" a region $X$, if $r\cap X\neq \emptyset$; a rod $r$
``{\em is contained in}" a region $X$, if $r\cap X^c=\emptyset$, in which case we write $r\subseteq X$.

The rod configurations in $\O_X$ can contain overlapping and even 
coinciding rods; we denote by $R(r)$ the multiplicity of $r$ in $R\in\O_X$.
The grand canonical partition function in $X$ with open boundary conditions is
\be 
Z_0(X) = \sum_{R\in \Omega_{X}} z^{|R|} \varphi (R)
\ee
where $|R|:=\sum_{r} R(r)$ and $\varphi (R) $ implements the hard core interaction:
\begin{equation}\label{phi}
\varphi (R)= \prod_{r,r'\in R} \varphi (r,r'),\qquad  \varphi (r,r') =
\left\{
\begin{array}{ll}
1 & \mbox{if}\ r\cap r'=\emptyset \\
0 & \mbox{if}\ r\cap r'\neq\emptyset. \\
\end{array} \right.
\end{equation}
Let $\ell:=\lceil k/2 \rceil$ and assume that $\L\subseteq\ZZZ^2$ is a square box of side divisible by 
$4\ell$. We pave $\L$ by squares of side $\ell$, called 
``{\em tiles}", and by squares of side $4\ell$, called ``{\em smoothing squares}". The lattice of the tiles'  
centers is a  coarse grained lattice of mesh $\ell$, called $\L'$; similarly, the lattice of the smoothing squares'  
centers is a coarse grained lattice of mesh $4\ell$, called $\L''$.
Given $\x\in\L'$, the tile centered at $\x$ is denoted by $\D_\x$; given $a\in\L''$, 
the smoothing square centered at $a$ is denoted by $\SS_a$.
Given two sets $X,Y\subseteq \L$, we indicate their euclidean distance by $\dist(X,Y)=
\min_{x\in X, y\in Y}|x-y|$. If $X$ and $Y$ are union of tiles, 
we shall also indicate by $X',Y'\subset \L'$ the coarse versions of $X$ and $Y$, i.e., the sets of sites in 
$\L'$ such that $X=\cup_{\x\in X'}\D_\x$ and $Y=\cup_{\x\in Y'}\D_\x$. The distance 
between $X'$ and $Y'$ is denoted by $\dist(X',Y')$ and their rescaled 
distance by $\dist'(X',Y'):=\ell^{-1}\dist(X',Y')$; with these conventions, if $\x$ and $\h$ 
are nearest neighbor sites on $\L'$, then $\dist(\x,\h)=|\x-\h|=\ell$ and $\dist'(\x,\h)=1$.
The complement of $\L$ is denoted by $\L_c:=\ZZZ^2\setminus \L$ and its coarse version by $\L_c'$,
with obvious meaning.

The size of the tiles is small enough to ensure that if one vertical 
(horizontal) rod belongs to a given tile, then all other rods belonging to the same tile and respecting
the hard core repulsion condition must be vertical (horizontal). If a tile is empty, i.e., no rod 
belongs to it, then we assign it an extra fictitious label, which can take three possible values, 
either  $0$ or $+$ or $-$. 
A rod configuration $R\in \Omega_{\Lambda}$ (combined with an assignment of these extra 
fictitious labels) induces a spin configuration $\s=\{\s_\x\}_{\x\in\L'}$
on $\Lambda '$, $\s_\x\in\{-1,0,+1\}$, via the following rules: 
\begin{itemize}
\item[-] $\sigma_{\xi }=+1$, if all rods belonging to $\D_\xi $ are horizontal or if the tile is empty with the extra label equal to $+$,
\item[-] $\sigma_{\xi }=-1$, if all rods belonging to $\D_\xi $ are vertical or if the tile is empty
with the extra label equal to $-$,
\item [-]  $\sigma_{\xi }=0$, if $\D_\x$ is empty with the extra label equal to $0$.
\end{itemize}
The corresponding set of rod configurations in the tile $\D_\x$ is denoted by 
$\O^{\s_\x}_{\D_\x}$: 
$\Omega^{+}_{\D_\x}$ ($\Omega^{-}_{\D_\x}$) 
is the set of rod configurations in $\Delta_{\xi}$ consisting either of horizontal 
(vertical) rods or of the empty configuration; similarly, $\Omega^{0}_{\D_\xi }$ consists
only of the empty configuration. 

Note that the grand canonical partition function in $\L$ with open boundary conditions
can be rewritten as
\be Z_0 (\Lambda )= \sum_{ \s\in \Theta_{\L'}} \sum_{R\in\O_\L(\s)}
\bar\varphi (R)\;,\label{3.3}\ee
where $ \Theta_{\L'}:=\{-1,0,+1\}^{\L'}$ and $\O_\L(\s):=\cup_{\x\in\L'}\O^{\s_\x}_{\D_\x}$. Moreover,
\be \bar\varphi (R):=\Big[\prod_{\x\in\L'}\z(\x)\Big]\f(R)\;,\label{3.4}\ee
where the activity of a tile is defined as
\be\zeta (\xi) =\left\{
\begin{array}{ll}
z^{|R_{\xi }|} & \mbox{if}\  \sigma_{\xi }=\pm 1\\
-1  & \mbox{if}\  \sigma_{\xi }=0.\\
\end{array}\right.\ee
The sign $-1$ is necessary to avoid over-counting of the empty configurations. Note that
$\bar\f(R)$ depends both on $\s$ and on $R$; however, in order not to overwhelm the notation, we 
shall drop the label $\s$.

The partition function with $q$ boundary conditions, $q=\pm$, denoted by $Z(\L|q)$,
can be defined in a similar fashion:
\be Z(\L|q)= \sum_{\s\in\Theta^q_{\L'}} 
 \sum_{R\in\O_\L(\s)}\bar\varphi (R)\label{3.6}\ee
where $\Theta_{\L'}^q\subset\Theta_{\L'}$ is the set of spin configurations such that 
$\dist'(\x,\L_c')\le 5\Rightarrow\s_\x=q$. The number 5 appearing here is related to the
choice of smoothing squares of side $4\ell$: in fact, the condition that all the spins $\s_\x$ with 
$\dist'(\x,\L_c')\le 5$ are equal to $q$ guarantees that all the smoothing squares adjacent to the boundary 
of $\L$ are uniformly ``magnetized" with magnetization $q$ and that, moreover, all such smoothing squares are surrounded by a $1$-tile-thick peel of spins equal to $q$. These two conditions are convenient for an explicit construction of a contour representation for $Z(\L|q)$, as we will show below. 
 
Correspondingly, the ensemble $\media{\cdot}_\L^{q}$ with $q$ boundary conditions
is defined by
\be \media{A_X}_\L^{q}=\frac1{Z(\L|q)}
 \sum_{\s\in\Theta^q_{\L'}} 
 \sum_{R\in\O_\L(\s)}\bar\varphi (R)\, A_X(R)\;,\label{3.7}\ee
where $A_X$ is a local observable, depending only on the restriction $R_X$ of the rod 
configuration $R$ to a given finite subset $X\subset\L$. The infinite volume states $\media{\cdot}^q$ with $q$ boundary conditions are defined by 
\be \media{A_X}^{q}=\lim_{|\L|\to\infty}\media{A_X}_\L^{q}\;,\label{3.8}\ee
if the limit exists for all local observables $A_X$, $X\subset\ZZZ^2$. 
Our main results can be stated as follows.

\begin{theor}\label{thm1}
If $zk$ and $(zk^2)^{-1}$ are small enough, then 
the two infinite volume states $\media{\cdot}^q$, $q=\pm$, exist. They are translationally invariant 
and are different among each other. In particular, if $\c^\s_{\x_0}$ is the projection onto the rod 
configurations such that $R_{\x_0}\in\O_{\D_{\x_0}}^\s$, then 
\be \media{\c^{-q}_{\x_0}}^q\le e^{-c z k^{2}}\;,\label{3.9}\ee
for a suitable constant $c$. Moreover, let $n_{x_0}$ be the indicator function that is equal to 
$1$ if a rod has a center in $x_0\in\ZZZ^2$ and $0$ otherwise, then
\be \r=\media{n_{x_0}}^+=\media{n_{x_0}}^-=z(1+O(zk,e^{-c zk^2}))\label{3.10}\ee
and
\be \r(x-y)=\media{n_{x}n_{y}}^+=\media{n_{x}n_{y}}^-=
\r^2\Big(1+O(e^{-c|x-y|/k})\Big)\;,\label{3.11}\ee
for a suitable $c>0$. \end{theor}

Eq.(\ref{3.9}) proves the existence of orientational order in the system. Eqs.(\ref{3.10})-(\ref{3.11})
prove the absence of translational symmetry breaking. These two behavior together prove that 
the system is in a nematic liquid crystal phase, as announced in the introduction. 
The rest of the paper is devoted to the proof of Theorem \ref{thm1}, which is based on 
a two-scales cluster expansion.  As it will be clear from the discussion in the next sections, 
our construction proves much more than what is explicitly stated in Theorem \ref{thm1}, namely it 
allows us to compute the averages of all the local observables in terms of an explicit exponentially convergent series. 


\section{The contour theory.}\label{sec4}
\setcounter{equation}{0}
\renewcommand{\theequation}{\ref{sec4}.\arabic{equation}} 


The proof of the Theorem \ref{thm1} will be split in several steps. We start by developing a 
representation of the partition function $Z(\L|q)$ with $q$ boundary conditions in terms
of a set of interacting contours. Later, we will adapt the contour expansion to the computation 
of the correlations. The contour theory can be studied by an adaptation of Pirogov-Sinai's method
to the present context. See \cite{PS} for the original version of this method and \cite{BI, K, Z, Z2} 
for several alternative simplified versions of it. In the following we will try to be as self-consistent as
possible and to keep things simple, by avoiding as much as we can general and abstract settings.
We first need some more definitions.\vskip.2truecm

\begin{figure}[t]
\centering
\includegraphics[width=0.8\textwidth]{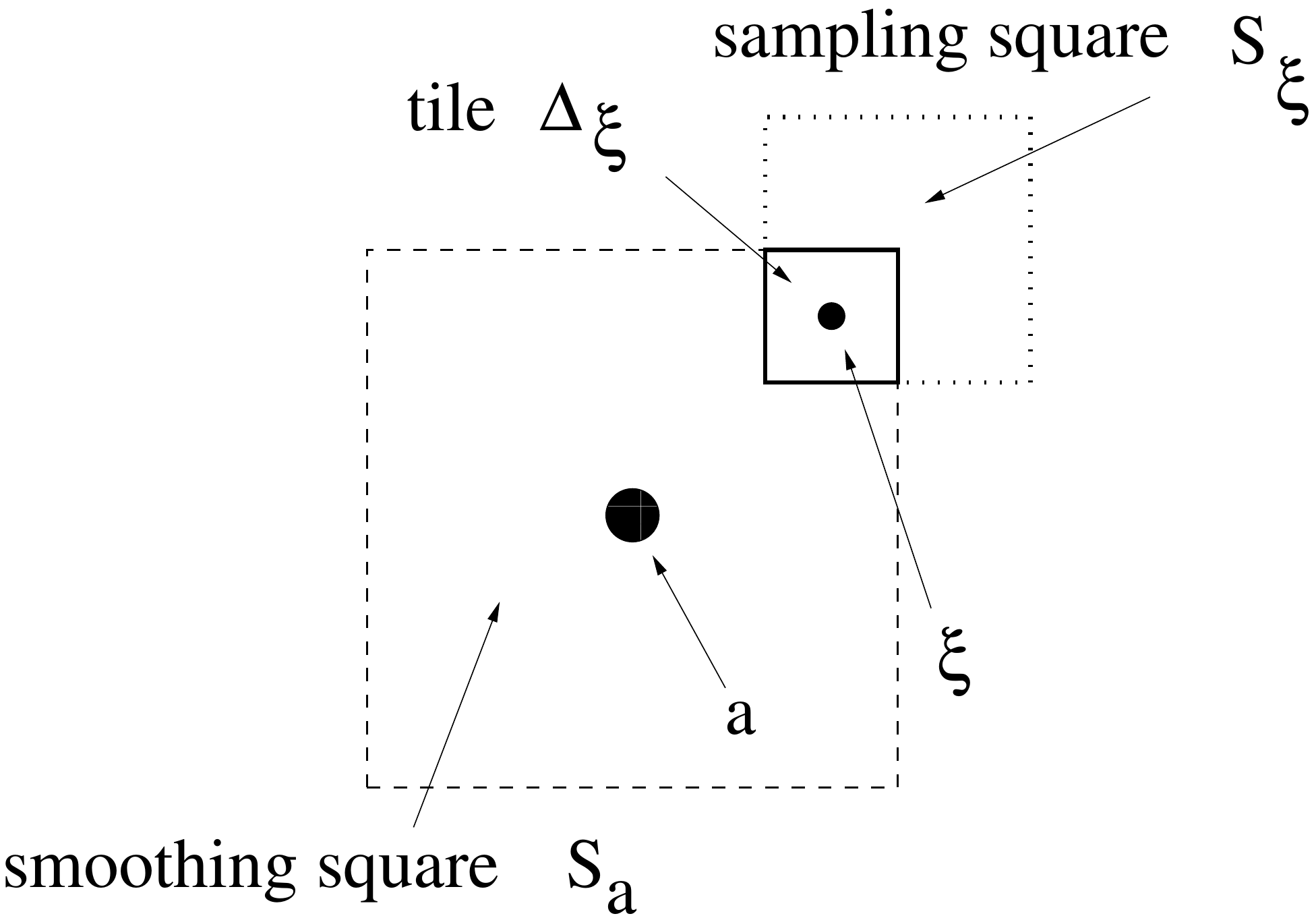}
\caption{An example of tile, sampling square and smoothing square}
\label{fig1}
\end{figure}

{\bf Definition 1:  sampling squares.} Given a spin configuration 
$\s\in\Theta^q_{\L'}$, this induces a partition of
$\Lambda'$ into regions where the spins are ``uniformly magnetized up or down" (i.e., 
regions where the spins are constantly equal to $+1$ or to $-1$) and boundary regions 
separating the ``uniformly magnetized regions" among each other, which can possibly contain 
spins equal to zero. To make this more precise we introduce the notion of 
``sampling squares", defined as follows: given $\x\in\L'$, the sampling 
square associated to $\x$ is defined as $S_\x= \cup_{\h\in\L'\,:\  0\le \h_i-\x_i\le \ell}\, \D_\h$, 
where $\h_i$ and $\h_i$, $i=1,2$, are the coordinates of $\x,\h\in\L'$.
Note that if $\dist'(\x,\L_c')>1$, then $S_\x$ 
contains exactly {\em 4 tiles}. See Fig.\ref{fig1} for an example.  We say that a sampling square is 
\begin{itemize}
\item{\bf good} if the spins inside $S_\x$ are all equal either to $+1$ or to $-1$. 
Each good sampling square comes with a {\it magnetization} $m=\pm 1$. 
\item{\bf bad} otherwise; note that each bad sampling square is such that either 
it contains at least one spin equal to zero, or it contains at least one pair of neighboring spins with
opposite values, $+1$ and $-1$.
\end{itemize}

{\bf Definition 2: connectedness, good and bad regions.} 
Given a configuration $\s\in\Theta_{\L'}$, we call
\be B (\sigma )= \cup_{\substack{\hskip-.3truecm\xi \in \Lambda':\\ S_\x\, {\rm is}\, {\rm bad} }} S_\x\label{4.1}\ee
the union of all bad sampling squares. The ``smoothening" of $B(\s)$ on scale $4\ell$ is defined as:
\be \lis B(\s)=\cup_{\substack{\hskip-.5truecm a\in\L'':\\ \SS_a\cap B(\s)\neq \emptyset}} \SS_a\;,\label{4.1aa}\ee
where the lattice $\L''$ and the smoothing squares $\SS_a$ were defined in the paragraph following  equation
(\ref{phi}).

Let $X\subseteq \L$ be a union of tiles: we say that $X$ is connected if, given any pair of points 
$x,y\in X$, there exists a sequence $(x_0=x,x_1,\ldots,x_{n-1},x_n=y)$ such that $x_i\in X$ and 
$|x_i-x_{i-1}|=1$, for all $i=1,\ldots,n$. We also say that $X$ is D-connected 
(with the prefix ``D" meaning ``diagonal") if, given any pair of points $x,y\in X$, there exists a 
sequence $(x_0=x,x_1,\ldots,x_{n-1},x_n=y)$ such that $x_i\in X$ and $|x_i-x_{i-1}|\le \sqrt2$, for 
all $i=1,\ldots,n$ (here $|x-y|$ is the euclidean distance between $x$ and $y$). 

The maximal D-connected components of $\lis B(\s)$ are denoted by $\G_j$ and are the geometric
supports of the contours that we will introduce below. The complement of the bad region, 
\be G(\s):=\L\setminus \lis B(\s)\;,\label{4.2}\ee
can be split into uniformly magnetized disconnected regions, each of which is a union of tiles; 
these are denoted by $Y_j$ and $m_j$ are the corresponding magnetizations. 
\vskip.2truecm
\noindent{\bf Remarks.}
\begin{enumerate}\item Note that distinct D-disconnected bad regions in ${\lis B} (\sigma )$, 
$\G_j(\s), \G_{j'}(\s)$ with $j\neq j'$,
do not interact directly; i.e., $\f(R_\x,R_{\h})=1$ for all $\x\in \G_j$, $\h\in \G_{j'}$.
This is because $\G_j(\s)$ and $\G_{j'}(\s)$ are separated by at least one smoothing square
(hence 4 tiles). Similarly, 
distinct uniformly magnetized 
disconnected regions, $Y_j(\s), Y_{j'}(\s)\in G (\sigma )$ with $j\neq j'$ and magnetizations $m_j$, $m_{j'}$,
do not interact directly; i.e., $\f(R_\x,R_{\h})=1$ for all 
$\x\in Y_j$, $\h\in Y_{j'}$ and for all $R_\x\in \O^{m_j}_\x$, $R_{\h}\in \O^{m_{j'}}_{\h}$.
In fact, note that 
$R_\x$ and $R_\h$ can interact only in one
of the following two cases: $\x$ and $\h$ are on the same
row (column) and $|\x-\h|\le 2\ell$, or $|\x_1-\h_1|=|\x_2-\h_2|=\ell$.
If $\x\in Y_j$ and $\h\in Y_{j'}$ with $j\neq j'$, then
the first case can occur only if $|\x-\h|\ge 5\ell$ (in the horizontal or vertical directions, 
$Y_j$ and $Y_{j'}$ are separated by at least one smoothing square), in which case 
$R_\x$ and $R_\h$ certainly do not 
interact, whatever is the alignment 
of the rods. In the second case necessarily 
$m_j=m_{j'}$, otherwise the sampling square containing both $\x$ and $\h$
would be bad and both tiles would belong to
$B (\sigma )$ instead of $G (\sigma )$. Now, if $m_j=m_{j'}$
the rods in $R_\x$ have the same orientation as those in $R_{\h}$, while their 
centers belong to different rows and columns and, therefore, do not interact. 

\item In terms of the definitions above, the set $\Theta_{\L'}^q\subset \Theta_{\L'}$ of spin
configurations with $q$ boundary conditions can be thought as the set of spin configurations 
such that all the contours' supports $\G_j\subset \lis B(\s)$ are D-disconnected from $\L^c$
and separated from it by at least one smoothing square. 
\end{enumerate}

\begin{figure}[t]
\centering
\includegraphics[width=0.5\textwidth]{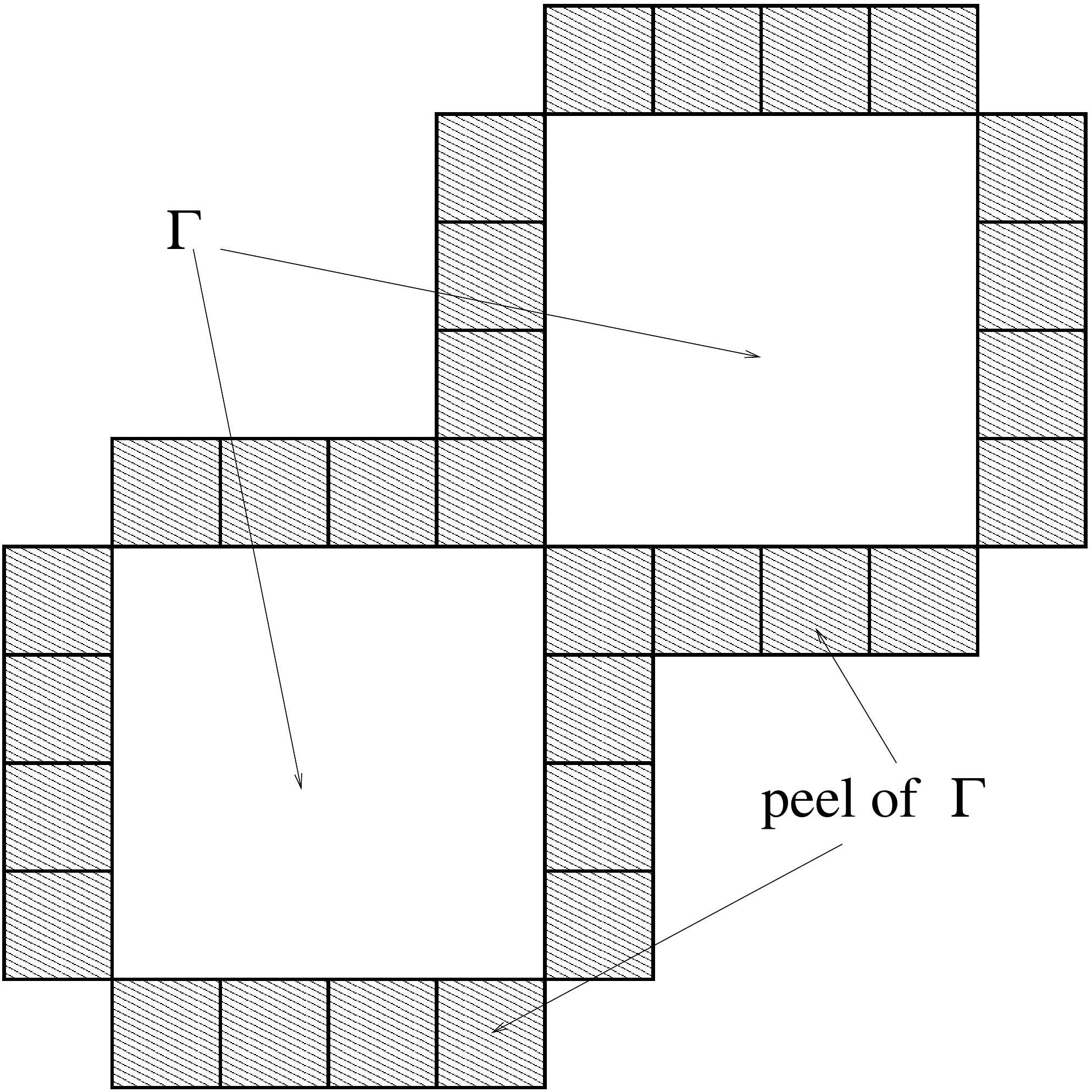}
\caption{The peel of a bad region $\Gamma$}
\label{fig2}
\end{figure}

\vskip.2truecm
{\bf Definition 3: contours.} 
Given a spin configuration with $q$ boundary conditions $\s\in\Theta_{\L'}^q$ and 
a rod configuration $R\in\O_\L$ compatible with it,
let $\G$ be one of the maximal connected  components of $\lis B(\s)$. 
By construction, the complement of $\G$, $\L\setminus \G$, consists of one or more connected 
components: one of these components is adjacent to (i.e., it is at a distance $1$ from) $\L^c$ 
and is naturally identified as the exterior of 
$\G$; it is denoted by ${\rm Ext}\,\G$. If $\G$ is simply connected this is the only connected 
component of $\L\setminus \G$;
if not, i.e., if $\G$ has $h_\G\ge 1$ holes, then there are other connected components of 
$\L\setminus \G$, to be called the interiors of $\G$ and denoted by ${\rm Int}_j\G$, $j=1,\ldots,
h_\G$. The interior of $\Gamma $ is then ${\rm Int}\Gamma = \cup_{j}{\rm Int}_j\G$. 
For what follows, it is also convenient to introduce the {\em 1-tile-thick peel } of $\G$ (see Fig.\ref{fig2}): 
\be P_\G=\cup_{\substack{\hskip-.3truecm\x\in\L':\\ 
\dist'(\x,\G')=1}}\D_\x\;.\label{e4.4_0}\ee
Note that, since distinct $D$-disconnected regions are separated 
by at least one smoothing square (i.e., 4 tiles), then also the peels associated to distinct $\G$'s are mutually 
$D$-disconnected.

The {\em contour} $\gamma$ associated to the support $\G=\supp(\g)$ is defined as the collection:
\be 
\gamma = (\Gamma , \sigma_{\gamma }, R_{\gamma }, m_{ext}, \underline{m}_{int})
\label{e4.4}\ee
where
\begin{itemize}
\item  $\sigma_{\gamma }$ is the restriction of the spin configuration $\s$ to $\G$;
\item $R_\g$ is the restriction of the rod configuration $R$ to $\G$;
\item  $m_{ext}$ is the magnetization of $P_\G^{ext}:={\rm Ext}\,\G\cap P_\G$;
\item $ \underline{m}_{int}=\{m_{int}^1,\ldots,m_{int}^{h_\G}\}$, with $m^j_{int}$
the magnetization of ${\rm Int}_j\G\cap P_\G$; if $h_\G=0$, then $ \underline{m}_{int}$ is the empty set.
In the following we shall also denote by $P_\G^{int}:=\Int\G\cap P_\G$ the internal peel of $\G$. 
\end{itemize}
If $m_{ext}=q$, then we say that $\g$ is a $q$-contour. 
\vskip.2truecm
\noindent{\bf Remark.} The set $\g$ must satisfy a number of constraints. In particular, given 
$m_{ext}$ and $\ul m_{int}$, $\s_\g$ must be compatible with the conditions that: (i) all the 
sampling squares having non-zero intersection with $P_\G$ are good (otherwise the contour would also contain these squares); (ii) each smoothing square contained in $\G$ has non 
zero intersection with at least one bad sampling square. 
Moreover, $R_\g$ must be compatible with $\s_\g$ itself.
\vskip.2truecm
In the following we want to write an expression for  $Z (\Lambda|q )$ purely in terms of contours. 
Roughly speaking, given a contour configuration contributing to the r.h.s. of Eq.(\ref{3.6}), we first 
want to freeze the rods inside the supports of the contours, next sum over all the rod configurations
in the good regions and show that the resulting effective theory is a contour theory treatable
by the Pirogov-Sinai method. The resummation of the configurations within the good regions can 
be performed by standard cluster expansion methods, as explained in the following digression.
\vskip.2truecm
{\bf Partition function restricted to a good region.} Given a set $X\subseteq\L$ consisting of a union of tiles,
let $\O^q_{X}=\cup_{\x\in X'}\O^q_{\D_\x}$, $q=\pm$.
The restricted theory of the ``uniformly $q$-magnetized" region $X$ (with open boundary 
conditions) is associated to the partition function:
\be Z^{q}(X) = \sum_{R\in \Omega_{X}^{q}}z^{|R|}{\varphi } (R)\;,\label{4.4}\ee
which can be easily computed by standard cluster expansion methods, some aspects of which are briefly reviewed here (for extensive reviews, see, e.g., \cite{Bry} and \cite[Chapt. 7]{GBG}). 
The logarithm of Eq.(\ref{4.4}) can be expressed in terms of a convergent series as:
\be \log Z^{q}(X) = 
\sum_{R\in \Omega_{X}^{q} }z^{|R|}{\varphi}^{T} (R)  =
z|X| (1+O (zk))\label{04.3}\ee
where $\f^T$ are the {\it Mayer's coefficients}, which admit the following explicit representation.
Given the rod configuration $R=\{r_1,\ldots,r_n\}$, consider
the graph $\GG$ with $n$ nodes, labelled by $1,\ldots,n$, with edges
connecting all pairs $i,j$ such that $r_i\cap r_j\neq\emptyset$ ($\GG$ is sometimes called the 
connectivity graph of $R$). Then one has $\Ft^T(\emptyset)=0$, $\Ft^T(r)=1$ and, for $|R|>1$:
\be \Ft^T(R)=\fra{1}{R!} \sum_{C\subseteq \GG}^* (-1)^{\hbox{\nota number of
edges in}\ C} , \label{04.4}\ee
where $R!=\prod_{r} R(r)!$ and the sum runs over all the connected 
subgraphs $C$ of $G$ {\it that
visit all the $n$ points} $1,\ldots,n$. In particular, if $|R|>1$, then $\f^T(R)=0$ unless $R$ is connected.

The sum in the r.h.s. of Eq.(\ref{04.3}) is exponentially convergent 
for $zk\ll1$; in 
particular, if $x_0\in\L$, then for a suitable constant $C>0$
\be \sum_{\substack{R\in \O^q_{\L}:\\
R\ni x_0,\ |R|\ge m}}|z|^{|R|}|\f^T(R)|\le Cz(Czk)^{m-1}\;,\label{R4.9}\ee
uniformly in $\L$, where $R\ni x_0$ means that $R$ contains at least one rod with center in $x_0$.
Moreover, the sum $\sum_{R\in \O^q_{\L}:\, R\ni x_0}z^{|R|}\f^T(R)$ is
analytic in $zk$, uniformly in $\L$, for $zk$ small enough and its limit as $\L\nearrow 
\ZZZ^2$ is analytic, too. A useful corollary of Eq.(\ref{R4.9}) is the following: 
if $V(R)$ is the union of the centers of the rods in $R$, 
${\rm supp} (R)$ is the support of the union of rods $r\in R$ (thought of as a subset of $\L$)
and $\diam(\supp(R))$ is its diameter, then for any finite 
region $X\subset \Lambda$:
\bea 
\sum_{\substack{R\in \O^q_{\L}:\\
V (R)\cap  X\neq \emptyset,\\
 {\rm diam}({\rm supp}(R) )\ge d}}|z|^{|R|}|\f^T(R)|&\le& 
\sum_{x_0\in X}\sum_{m\ge \lceil \frac{d}{k-1}\rceil  } \sum_{\substack{R\in \O^q_{\L}:\\
R\ni x_0,\ |R|\ge m}}|z|^{|R|}|\f^T(R)|\nn\\
&\le &  2Cz|X|(Czk)^{\frac{d}{k-1}-1}\;,
\label{R4.9bis}\eea
uniformly in $\L$; here, in the first inequality, we used the fact that in order for $\supp(R)$ to have diameter $d$, the 
configuration $R$ needs to have at least $\lceil d/(k-1)\rceil$ rods, while in the second inequality we 
used  Eq.(\ref{R4.9}). 
In a similar fashion, all the correlation functions can be computed 
in terms of convergent series, as long as $zk$ is small enough. 
These results are classical, see \cite{Ru} or, e.g., \cite{Bry,GBG}.
The restricted theory is applied to the computation of the sums over the rod configurations in
the good regions, as described in the following.
\vskip.2truecm
{\bf Contour representation of the partition function.}
Given a contour $\g$, let $Z_{\gamma } (\Int_j\G|m_{int}^j)$ be the partition function 
on the $j$-th interior of $\Gamma$ with the boundary conditions created by 
the presence of the ``frozen" rods $R_\g$. Moreover, 
if  $\x\in P_\G'$, let 
\be A_\g(\D_\x)=\cup_{\h\in a_\g(\x)}\D_\h\;, \qquad C_\g(\D_\x)=\cup_{\substack{\hskip-.5truecm
\h\in \G':\\ \dist'(\h,\x)\le 2}}\D_\h\;,\label{c4.16a}\ee
where
\be a_\g(\x):=\{\x\}\cup\{\h\in\L': \dist'(\h,\x)=1, \dist'_1(\h,\G')=2, \h_{j(-q)}=\x_{j(-q)}\}
\;,\label{c4.16}\ee
with $\dist'_1(\cdot,\cdot)$ the rescaled (``coarse") $L_1$ distance on $\L'$ and $j(+)=1$, $j(-)=2$.

\begin{figure}[t]
\centering
\includegraphics[width=0.6\textwidth]{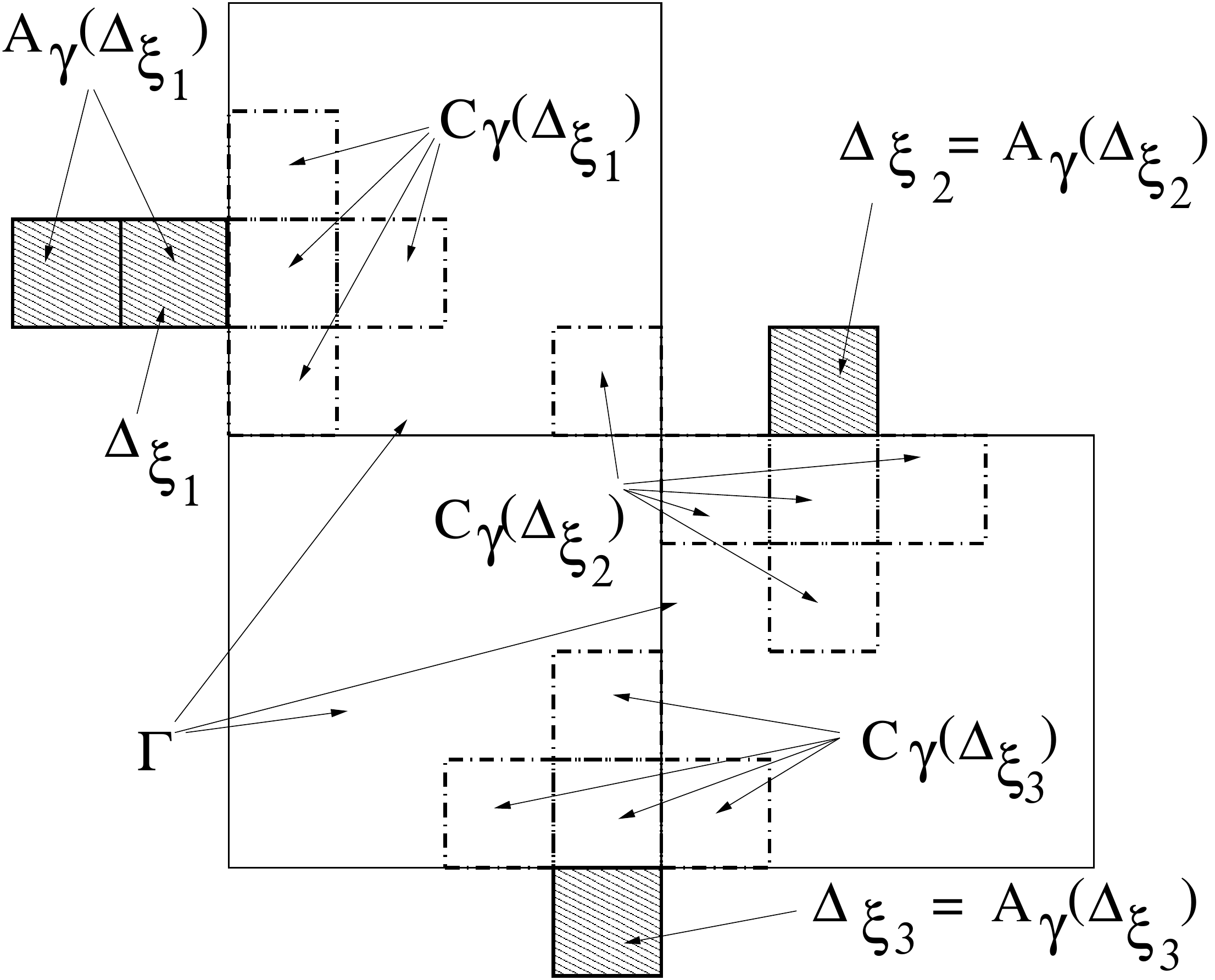}
\caption{The two sets $A_\g(\Delta_{\xi})$ and  $C_\g(\Delta_{\xi})$ in the case that $q=+$.}
\label{fig3}
\end{figure}

Finally, given $\D\subseteq P_\G$, let $f_\D$ and $g_\D$ be the following characteristic functions:
\begin{align} 
f_{\D} (R) &= 
\left\{
\begin{array}{ll}
1 & \mbox{if $R$ has at least one rod belonging to $A_\g(\D)$}\\
&\mbox{and one belonging to $C_\g(\D)$}\;, \label{eq:f}\\
0& \mbox{otherwise}\;,\\
\end{array}\right.\\
g_{\D} (R) &=\label{eq:g}
\left\{
\begin{array}{ll}
1 & \mbox{if $R\cap R_\g\neq\emptyset$, $R$ has at least one rod belonging to $A_\g(\D)$}\\
&\mbox{and 
$R_\g$ has at least one rod belonging $C_\g(\D)$}\;, \\
0& \mbox{otherwise}\;.\end{array}\right.\end{align}
Pictorially speaking, $f_\D$ is the characteristic function of the event ``$R$ crosses the boundary of 
$\G$ at $\D$", while $g_\D$ is the characteristic function of the event ``$R$ intersects $R_\g$ across
$\D$".  Note that, by construction, given two distinct tiles, $\D_1\subseteq P_{\G_1}$ and 
$\D_2\subseteq P_{\G_2}$ such that $\D_1\cap\D_2=\emptyset$, then $A_{\g_1}(\D_1)\cap A_{\g_2}(\D_2)=\emptyset$, even in the case that $\G_1\equiv\G_2$.

In terms of these definitions, the following contours' representation for $Z(\L|q)$ is valid.
\begin{lemma}\label{lemma:contours}
The conditioned partition function $Z (\Lambda|q )$, $q=\pm 1$, can be written as
\be 
Z (\Lambda|q )= Z^{q} (\Lambda )\ 
\sum_{\dpr  \in \CC(\Lambda,q) }   
\Big[\prod_{\gamma \in\partial   } \zeta_{q} (\gamma ) \Big] 
\ e^{-W (\dpr)}\;, \label{lemma2}\ee
where:\begin{itemize}
\item $\CC (\Lambda,q )$ is the set of all the well D-disconnected $q$-contour 
configurations in $\Lambda$ (here we say that $\{\g_1,\ldots,\g_n\}$ is well D-disconnected
if the supports $\G_1,\ldots,\G_n$ are separated among each other and from $\L^c$ by at least one smoothing square); 
\item $\zeta_{q} (\gamma )$ is  the activity of $\gamma$:
\begin{equation}
\zeta_{q} (\gamma) =  \z^0_q(\g)\,\exp\Big\{-\sum_{R\in \O^q_\L}\f^T(R)z^{|R|}
\sum_{\D\subseteq P_\G}F_\D(R)\Big\}\;,\label{eq:zetadef}
\end{equation} 
where 
\be \z^0_q(\g)=\frac{\bar\varphi (R_{\gamma })  }{Z^{q}(\Gamma ) } \prod_{j=1}^{h_\G}  
\frac{ Z_{\gamma } (\Int_{j}\G |m^j_{int})}{ Z (\Int_j\G |q )}\label{eqz0}\ee
and $F_\D=f_\D$ if $\D\subseteq P^{int}_\G$ while $F_\D=f_\D+g_\D(1-f_\D)$ if $\D\subseteq P^{ext}_\G$.
\item $W(\dpr)$ is the interaction between the contours in $\dpr$:
\begin{equation}\label{4.20}
W (\dpr) = \sum_{R\in\O^q_\L}\f^T(R)z^{|R|}
\sum_{n\ge 2}(-1)^{n+1}\sum^*_{\D_1<\cdots<\D_n}F_{\D_1}(R)\cdots F_{\D_n}(R)\;,\ee
where the $*$ on the sum indicates the constraint that
$\D_1,\ldots,\D_n$ are all contained in the peel of some contour of $\dpr$ and their centers $\x_1,\ldots 
\x_n$ all belong to the same row (if $q=+$) or column (if $q=-$) of $\L'$, namely
$\x_{1,j(-q)}=\cdots=\x_{n,j(-q)}$. Moreover, by writing $\D_1<\cdots<\D_n$, we mean that 
$\x_{1,j(q)}<\cdots<\x_{n,j(q)}$. Finally, $F_\D=f_\D$ if $\D$ is contained in the internal peel of 
some contour in $\dpr$ or $F_\D=f_\D+g_\D(1-f_\D)$ if $\D$ is contained in the external peel of 
some contour in $\dpr$.
\end{itemize}
\end{lemma}

{\bf Remarks.}
\begin{enumerate}
\item 
The contour configurations $\{\g_1,\ldots,\g_n\}\in\CC(\L,q)$ consist of $n$-ples of 
well $D$-disconnected $q$-contours, which means that the geometric supports $\G_1,\ldots,\G_n$ are 
separated among each other and from $\L^c$ by at least one smoothing square. Note, however, 
that their external and internal magnetizations are not necessarily compatible among each other:
for instance, $\G_1$ may have one hole surrounding $\G_2$, and the internal magnetization 
of $\G_1$ may be different from the external magnetization of $\G_2$ (which is $q$). It is 
actually an important point of the representation Eq.(\ref{lemma2}) that we can forget about 
the compatibility conditions among the internal and external magnetizations of different contours.
There exist different (and even more straightforward)
contour representation of $Z(\L|q)$ where the internal and external contours' magnetizations satisfy
natural but non-trivial constraints (e.g., in the example above, the natural constraint is that the
internal magnetization of 
$\G_1$ is the same as the external magnetization of $\G_2$). However, the magnetization 
constraints are not suitable to apply cluster expansion methods to the resulting contour theory. 
Therefore, it is convenient to eliminate such constraints, at the price of adding the extra factors
$Z_{\gamma } (\Int_{j}\G |m^j_{int} ) / Z (\Int_j\G |q )$ in the definition 
of the contours' activities, 
see Eq.(\ref{eqz0}).
\item The interest of the representation Eq.(\ref{lemma2}) is that the contour activities and the 
multi-contour interaction satisfy suitable bounds, allowing us to study the r.h.s. of Eq.(\ref{lemma2})
by cluster expansion methods. In particular, 
 $\sup_{\s_\g}^*\sum_{R_\g\in\O_\G(\s_\g)}|\z_q(\g)|\le \exp\{-(\const.)zk^2|\G'|\}$,
where the $*$ on the sup reminds the constraint that all the smoothing squares in $\G$
must have a non-zero intersection with at least one bad sampling square. 
Moreover, $W(\dpr)$ is a quasi-one-dimensional potential, 
exponentially decaying to zero  in the mutual distance 
between the supports of the contours in $\dpr$. The proofs of these claims will be postponed to the 
next sections.
 \end{enumerate}

\vskip.2truecm
{\bf Proof of Lemma \ref{lemma:contours}.} 
Given $\s\in\Theta^q_{\L'}$ a spin configuration with $q$ boundary conditions 
consider the corresponding set of contours $\{\g_1,\ldots,\g_n\}$. 
Some of them are {\it external}, in the sense that they are not surrounded by any other contour in 
$\{\g_1,\ldots,\g_n\}$. 
By construction, these external contours are all $q$-contours. 
We denote by $\CC_{ext}(\L,q)$ the set of external 
$q$-contour configurations. Given $\dpr\in\CC_{ext}(\L,q)$, there is a common external 
region to all the contours in $\dpr$, which we denote by $\Ext(\dpr)$.
Besides this, there are several internal regions within each contour $\g\in\dpr$.
For each external contour $\g\in\dpr$, we freeze the corresponding rod configuration $R_\g$ and sum 
over the rod configurations inside all the internal regions $\Int_j\G$, $j=1,\ldots,{h_\G}$.
In this way, for each such interior, we reconstruct the partition function 
$ Z_{\gamma } (\mbox{Int}_{j}\G | m_{int}^j )$. On the other hand, by construction all rods inside 
$\mbox{Ext} (\partial)$ are either horizontal or vertical, according to the value of $q$.
Therefore, if we sum over all the allowed rod configurations inside this region we get the 
restricted partition function $Z^{q}_{\partial} (\mbox{Ext} (\partial) ) $, 
where the subscript $\partial$
reminds the fact that the rods $R_\dpr=\cup_{\g\in\dpr}$ create an excluded
volume for the rods in  $\Ext(\dpr)$. Using these definitions, we can rewrite  
\be 
Z (\Lambda|q )= \sum_{\partial\in\CC_{ext}(\Lambda,q)}  
Z^{q}_{\partial}(\mbox{Ext}(\partial))
\prod_{\gamma \in\partial}\Big[\bar\varphi (R_{\gamma })\prod_{j=1}^{h_\G}
Z_{\gamma } (\mbox{Int}_{j}\G | m_{int}^j)\Big]\;.
\label{4.17}\ee
Note that here we used the fact that the exterior and the interior(s) of $\dpr$ do not interact
directly (i.e., they only interact through $R_\g$). Using the definition of $\z^0_q(\g)$, 
Eq.(\ref{eqz0}), we can rewrite $Z(\L|q)$ as
\be
 \frac{ Z (\Lambda|q ) }{Z^{q}(\Lambda)} = \sum_{\partial\in\CC_{ext}(\Lambda,q)}  
\prod_{\gamma \in\partial} \Big[\zeta_{q}^{0} (\gamma)
\prod_{j=1}^{h_\G}\frac{Z (\mbox{Int}_{j}\G|q)}{Z^q(\mbox{Int}_{j}\G)}\Big]
e^{-W_0^{ext} (\dpr)}\;,
\label{4.177}\ee
where
\begin{equation}
e^{-W_{0}^{ext} (\dpr)}= 
\frac{ Z^{q}_{\partial} (\mbox{Ext} (\partial) )\prod_{\gamma \in\partial}
\big[  Z^{q}(\Gamma ) \prod_{j=1}^{h_\G} Z^{q} (\mbox{Int}_{j}\G)\big]}{Z^{q}(\Lambda) }.
\end{equation}
The factors $\frac{Z (\Int_{j}\G|q)}{Z^q(\Int_{j}\G)}$ have the same form as the l.h.s. 
of Eq.(\ref{4.177}) itself, with $\L$ replaced by $\Int_j\G$: therefore, the equation can be iterated 
until the interior of all the contours is so small that it cannot contain other contours. The result of the iteration is 
\be \frac{ Z (\Lambda|q ) }{Z^{q}(\Lambda)} = 
\sum_{\partial\in\CC(\Lambda,q)} \Big[\prod_{\gamma\in\partial} \zeta_{q}^{0}(\gamma) \Big]
\,e^{-W_{0} (\dpr)}\;,\label{4.19}\ee
where
\be 
e^{-W_{0} (\dpr)}=\frac{ Z^{q}_{\partial} (\Lambda (\partial) ) \prod_{\gamma \in\partial}
  Z^{q}(\Gamma ) 
}{ Z^{q}(\Lambda) },
\label{4.207}\ee
$\Lambda (\partial)= \Lambda \backslash \cup_{\gamma\in\dpr}\Gamma$ is the complement 
of the contours' supports and $Z^{q}_{\partial  } (\Lambda (\partial) )$ is the restricted partition 
function with magnetization $q$ in the volume $\L(\dpr)$ and in the presence of the hard rod 
constraint generated by the frozen rods $R_\g$ in the region 
$\cup_{\g\in\dpr}\cup_{\D\subseteq P^{ext}_\G}A_\g(\D)$. 

We now use Eq.(\ref{04.3}) and the analogous expression for $Z^q_\dpr(\L(\dpr))$, i.e., 
\be 
\log Z^{q}_{\partial} (\Lambda (\partial) )=
\sum_{R\in \Omega_{\L(\dpr)}^{q}}^{R{\buildrel ext\over \cap} R_\dpr=\emptyset}z^{|R|}{\varphi}^{T} (R)\;,
\label{4.21}\ee
where $R{\buildrel ext\over \cap} R_\dpr=\emptyset$ means that $R$ does not intersect $R_\dpr$ 
from the outside, namely:
\be R{\buildrel ext\over \cap} R_\dpr=\emptyset\quad{\buildrel def\over\Leftrightarrow}
\quad \prod_{\g\in\dpr}\prod_{\D\subseteq P^{ext}_\G}(1-g_\D(R))=1\;,\label{capext}\ee
where $g_\D$ was defined in Eq.(\ref{eq:g}). Then we can rewrite:
\begin{align}
e^{-W_{0} (\dpr)}=&
\frac{ Z^{q}(\Lambda (\partial) ) \prod_{\gamma \in\partial} Z^{q}(\Gamma ) }{ Z^{q}(\Lambda)}\cdot
\frac{ Z^{q}_{\partial  } (\Lambda (\partial) )  }{ Z^{q}(\Lambda (\partial) )}\,,
\label{4.22}\\
=&\exp\big\{-\sum_{R\in \Omega^{q}_\L}^{R{\buildrel \dpr\over\leadsto} 2}z^{|R|}\f^T(R)\big\}
\cdot\exp\big\{-\!\!\!\!\!\sum_{R\in \Omega^{q}_{\Lambda(\dpr)}}^{ R \cap R^{ext}_{\partial}\neq
\emptyset}z^{|R|}\f^T(R)\big\}
\, ,\nn
\end{align}
where $R{\buildrel \dpr\over\leadsto}2$ means that $R$ must contain two rods $r_1,r_2$ belonging,
respectively, to two distinct elements of
the partition $\PPP(\dpr)$ of $\L$ induced by the contours in $\dpr$; i.e., either $r_1,r_2\in R$ 
belong, respectively, to 
two disconnected components of $\Lambda (\partial)$, or they belong to two different contours' supports, or $r_1$ belongs to one 
contour's support and $r_2$ to one of the components of $\Lambda (\partial)$. 
Using the definitions of the characteristic functions $f_\D$ and $g_\D$ defined in 
Eqs.(\ref{eq:f})-(\ref{eq:g}), the two exponential in the r.h.s. of Eq.(\ref{4.22}) can be written as
\begin{align}
\sum_{R\in \Omega^{q} (\Lambda ) }^{R{\buildrel \dpr\over\leadsto} 2}z^{|R|}\f^T(R)&=
\sum_{R\in  \Omega^{q}_{\Lambda }} z^{|R|}\f^T(R)\Big[1-
\prod_{\g\in\dpr}\prod_{\D \subseteq P_\G}(1-f_\D(R))\Big]\;,
\label{4.23}\\
\sum_{R\in \Omega^{q}_{\Lambda(\dpr)}}^{ R \cap R^{ext}_{\partial}\neq
\emptyset}z^{|R|}\f^T(R) & =
\sum_{R\in \Omega^{q}_{\Lambda }} z^{|R|}\f^{T} (R)
\Big[ 1-
\prod_{\g\in\dpr}\prod_{\D \subseteq P^{ext}_\G}(1-g_\D(R))\Big]\cdot\label{4.25}\\
&\hskip2.95truecm\cdot\Big[
\prod_{\g\in\dpr}\prod_{\D \subseteq P_\G}(1-f_\D(R))\Big]
\;.\nn\end{align}
Using the representations Eqs.(\ref{4.22}), (\ref{4.23}), (\ref{4.25}) into Eq.(\ref{4.19}), we find
\bea
\frac{Z(\L|q)}{Z^q(\L)}&=&
\sum_{\partial\in\CC(\Lambda,q)} \Big[\prod_{\gamma\in\partial} \zeta_{q}^{0}(\gamma) \Big]
\exp\Big\{-\sum_{R\in  \Omega^{q}_{\Lambda }} z^{|R|}\f^T(R)\cdot\label{R2}\\
&&\cdot\Big[1-\Big(
\prod_{\g\in\dpr}\prod_{\D \subseteq P^{ext}_\G}(1-g_\D(R))\Big)\cdot\Big(
\prod_{\g\in\dpr}\prod_{\D \subseteq P_\G}(1-f_\D(R))\Big)\Big]\Big\}\;.\nn\eea
Note that the expression in square brackets in the second line can be conveniently rewritten as
\bea && 1-\prod_{\g\in\dpr}\Big(\prod_{\D \subseteq P^{ext}_\G}(1-g_\D(R))(1-f_{\D}(R))\Big)\cdot
\Big(\prod_{\D \subseteq P^{int}_\G}(1-f_{\D}(R))\Big)\equiv\nn\\
&&\equiv1-
\prod_{\g\in\dpr}\prod_{\D \subseteq P_\G}(1-F_\D)\;,\label{R3}\eea
where $F_\D$ was defined in the statement of Lemma \ref{lemma:contours}. Plugging Eq.(\ref{R3})
into Eq.(\ref{R2}) gives
\bea 
\frac{Z(\L|q)}{Z^q(\L)}&=&
\sum_{\partial\in\CC(\Lambda,q)} \Big[\prod_{\gamma\in\partial} \zeta_{q}^{0}(\gamma) \Big]
\exp\Big\{-\sum_{R\in  \Omega^{q}_{\Lambda }} z^{|R|}\f^T(R)\cdot\label{R4}\\
&&\cdot\Big[\sum_{\D\subseteq P_\dpr}F_\D(R) +\sum_{n\ge 2}(-1)^{n+1}\sum_{\{\D_1,\ldots,\D_n\}}
F_{\D_1}(R)\cdots F_{\D_n}(R)\Big]\;,\nn\eea
where the sum $\sum_{\{\D_1,\ldots,\D_n\}}$ runs over collections of distinct tiles 
$\D_i\subseteq P_\dpr$, with $P_{\dpr}:=\cup_{\g\in\dpr}P_{\G}$. Finally, using the fact that $\f^T(R)$
forces $R$ to be connected and, therefore, to live on a single row or column, depending on whether 
$q$ is $+$ or $-$, we find that the only non-vanishing contributions in the latter sum come from 
$n$-ples of tiles all living on the same row or column. This proves the desired result.
\qed


\section{Reorganizing the contour expansion.}\label{sec4bis}
\setcounter{equation}{0}
\renewcommand{\theequation}{\ref{sec4bis}.\arabic{equation}} 


Standard cluster expansion methods are more easily implemented in the case
of two-body interactions. Our contour interaction Eq.\eqref{4.20} is many-body but it can be reduced
to the two-body case by a slight reorganization of the expansion.

\begin{lemma}\label{lemma:contour-reorg}
The contour representation  (\ref{lemma2}) for the conditioned partition function $Z (\Lambda|q )$, $q=\pm 1$, can be reorganized as
follows
\be 
\frac{Z(\L|q)}{Z^q(\L)}= 1+\sum_{m\ge 1}\sum_{\{X_1,\ldots, X_m\}}
K_q^{(\L)}(X_1)\cdots K_q^{(\L)}(X_m)\, \phi(\{X_{1},X_{2}\dots ,X_{m}\})\;,
\label{5.4}\ee
where:
\begin{itemize}
\item each \emph{polymer} $X_{i}$ is a D-connected union of tiles in $\Lambda $;
\item  $\phi$ implements the hard core interaction, i.e., 
\bea &&\label{phipolymer}
\phi(\{X_{1},X_{2}\dots ,X_{m}\}) = \prod_{i<j} \phi (X_{i},X_{j}), \\ 
&&\phi(X_{i},X_{j})= \left\{ 
\begin{array}{ll}
1 & \mbox{\rm if  $X_{i}$  D-disconnected from $X_{j}$} \\
0&   \mbox{\rm otherwise.}\\
\end{array}
  \right.\nn\eea 
\item $ K_q^{(\L)}(X)$ is a suitable function of $X$, called the polymer's activity,
which is defined by Eq.(\ref{5.5}) below.  
\end{itemize}
\end{lemma}

\noindent {\bf Remark.}
The definition Eq.\eqref{phipolymer} of the polymer interaction is the analogue of Eq.\eqref{phi} 
with the rods replaced by polymers and the notion of intersection replaced by D-connectedness.

\begin{figure}[t]
\centering
\includegraphics[width=0.6\textwidth]{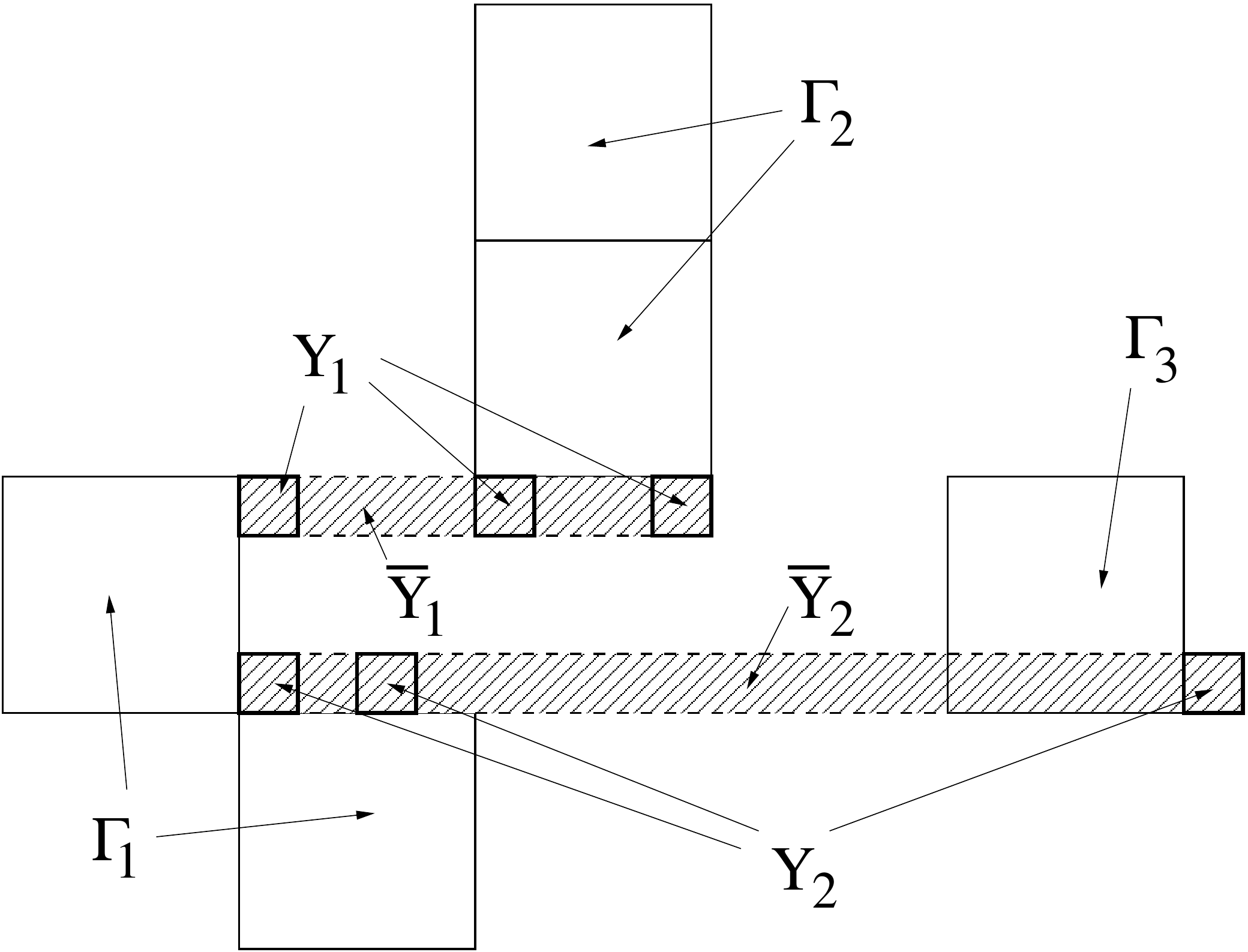}
\caption{An example of polymer $X$ and of a possible way of realizing it as a union of three 
contours' supports $\G_1,\G_2,\G_3$ and of two sets $Y_{1}$, $Y_{2}$.}
\label{fig4}
\end{figure}

\vskip.2truecm
 {\bf Definition of the polymer's activity.}
Given $\dpr=\{\gamma_{1},\dotsc \gamma_{n}\}\in\CC(\L,q)$ and $X_\dpr=\cup_{i=1}^n\G_i$,  let $Y=\{\Delta_{1},\dotsc \Delta_{m} \}$ be a 
collection of $m\geq 2$ 
distinct tiles, all contained in the peel of $X_\dpr$, i.e., $\Delta_{i}\in \cup_{j=1}^{n} P_{\Gamma_{j}}$,
and all belonging to the same row (if $q=+$) or column (if $q=-$). Since the tiles are all on the same 
row (column),
we can order them from left to right (bottom to top), $\Delta_{1}<\Delta_{2}<\dotsb <\Delta_{m}$. 
We denote by $\Y^q_{X_\dpr}$ the set of all such collections. 
Moreover, for each $Y=\{\Delta_{1},\dotsc \Delta_{m} \}\in\Y^q_{X_\dpr}$ with $\D_1<\cdots< \D_m$, we 
define $\lis Y$ to be the union of all the tiles between $\Delta_{1}$ and $\Delta_{m}$. 
With these definitions, the activity of the polymer $X$ is given by
\be 
K_q^{(\L)}(X)=\sum_{n\ge 1,\, p\ge 0}
\sum_{\substack{
\dpr\in\CC(\L,q): \ |\dpr|=n\\
\{Y_1,\ldots,Y_p\}:\ Y_i\in\Y^q_{X_\dpr}\\
X_\dpr\cup\{\cup_j\lis Y_j\}=X}}\Big[\prod_{\g\in\dpr}
\z_q(\g)\Big]\Big[\prod_{i=1}^p(e^{\FF(Y_i)}-1)\Big]\;,
\label{5.5}\ee
where  $\zeta_{q}(\gamma )$ was introduced in Eq.\eqref{eq:zetadef} and, if $Y=\{\D_1,\ldots,\D_m\}\in
\Y^q_{X_\dpr}$,
\be 
\FF(Y):=(-1)^{n}\sum_{R\in  \Omega^{q}_{\Lambda }} z^{|R|}\f^T(R)
F_{\D_1}(R)\cdots F_{\D_n}(R)\;.\label{5.2}\ee
An example of a polymer $X$ with non-vanishing activity and of a possible way of realizing it as 
a union of sets $\G_i$ and $\lis Y_j$ is given in Fig.\ref{fig4}.
\vskip.2truecm

{\bf Proof of Lemma \ref{lemma:contour-reorg}.} 
Using the definition of $\FF(Y)$, we can rewrite Eqs.(\ref{lemma2})-(\ref{4.20}) as:
\be
\frac{Z(\L|q)}{Z^q(\L)}=
\sum_{\dpr  \in \CC(\Lambda,q) }   
\Big[\prod_{\gamma \in\partial } \zeta_{q} (\gamma ) \Big] \Big[\prod_{Y\in\Y^q_{X_\dpr}}
 e^{\FF(Y)}\Big]\;.
\label{5.1}
\ee
Let us now add and subtract 1 to each of the factors $e^{\FF(Y)}$. In this way we
turn each factor into a binomial $1+(e^{\FF(Y)}-1)$. If $Y\in\Y^q_{X_\dpr}$, we associate the quantity 
$(e^{\FF(Y)}-1)$ with the region $\lis Y$; similarly, we associate the activity $\z(\g)$ with the region 
$\G$. In this way, every factor of the form $\prod_{i=1}^n\z(\g_i)\prod_{j=1}^p(e^{\FF(Y_j)}-1)$ 
is geometrically associated with the region $X=\{\cup_{i=1}^n\G_i\}\cup\{\cup_{j=1}^p\lis Y_j\}$. 
We develop the binomials $1+(e^{\FF(Y)}-1)$ and collect together the contribution 
corresponding to the maximally D-connected regions, obtained as unions of $\G_i$'s and $\lis Y_j$'s.  
The result is equation Eq.\eqref{5.4}. \qed 



\section{Convergence of the contours' expansion}\label{sec5}
\setcounter{equation}{0}
\renewcommand{\theequation}{\ref{sec5}.\arabic{equation}} 


In this and in the next section we prove the convergence of the cluster expansion 
for the logarithm of the partition function with $q$ boundary conditions, starting from 
Eq.(\ref{lemma2}). The proof will be split in two main steps: first, in this section, we prove convergence under the 
assumption that the activities $\z_q(\g)$ 
satisfy suitable decay bounds in the size of $|\G|$. Then, in the next section, we prove the validity of 
such a decay bound via an induction in the size 
of $|\G| $. From now on, $C,C',\ldots$ and $c,c',\ldots$
indicate universal positive constants (to be thought of as ``big" and ``small", respectively), whose 
specific values may change from line to line.

\begin{lemma}\label{lemma:convergence} Suppose that, for $zk$ and $(zk^2)^{-1}$ small enough, 
\begin{equation}\label{4.28}
\sup_{\s_\g}\!^*\sum_{R_\g\in\O_\G(\s_\g)}|\z_q(\g)|\le e^{-c_0\, zk^2|\G'|}\;,
\end{equation}
where the $*$ on the sup reminds the constraint that all the smoothing squares in $\G$
must have a non-zero intersection with at least one bad sampling square, and $c_0=5*10^{-4}$. 
Then the logarithm of the partition function admits a convergent cluster expansion 
\be 
\log Z(\L|q)=\sum_{R\in \Omega_{\L}^{q}}z^{|R|}{\varphi }^T (R)+
\sum_{{\cal X}\subseteq \L}\Big[\prod_{X\in{\cal X}}K^{(\L)}_q(X)\Big]
\phi^T({\cal X})\;,
\label{4.29}\ee
where ${\cal X}=\{X_1,\ldots,X_n\}$ is a polymers' configuration (possibly, some of the $X_i$'s may 
coincide), each polymer $X$ being a D-connected 
subset of $\L$ consisting of a union of 
tiles.
\end{lemma}

\noindent{\bf Remarks.} 
\begin{enumerate}
\item The constant $c_0=5*10^{-4}$ is a possible explicit constant for which the 
result of the lemma holds (certainly, it is not the sharp one). Its specific value is motivated 
by Lemma \ref{lemma:activities} and by its proof, see next section.  
\item The function $\phi^T({\cal X})$ in Eq.(\ref{4.29}) is the Mayer's coefficient of ${\cal X}$,
defined as in Eq.(\ref{04.4}), with $R$ replaced by ${\cal X}$, $r_i$ by $X_i$, and the notion ``$r_i\cap
r_j\neq\emptyset$" replaced by ``$X_i$ is  D-connected to $X_j$". 
\end{enumerate}

{\bf Proof.} By Lemma \ref{lemma:contour-reorg},   Eqs.(\ref{lemma2})-(\ref{4.20}) can be 
equivalently rewritten as
\be 
\frac{Z(\L|q)}{Z^q(\L)}= 1+\sum_{m\ge 1}\sum_{\{X_1,\ldots, X_m\}}
K_q^{(\L)}(X_1)\cdots K_q^{(\L)}(X_m)\, \phi(\{X_1,\dots X_{m}\})\;.\ee
It is well-known \cite{Ru,Bry,GBG}, that if the activities 
$K_q^{(\L)}(X)$ are sufficiently small and decay fast enough with the size of $X$, then one 
can apply standard cluster expansion methods (analogous to those sketched above, 
after Eq.(\ref{4.4}))  for computing the logarithm of Eq.(\ref{5.4}) and put it in the form of the exponentially 
convergent sum. More specifically, a sufficient  condition for the application 
of the standard cluster expansion is, see e.g. \cite[Proposition 7.1.1]{GBG},
\begin{equation}\label{polym-bound}
|K^{(\L)}_q(X)|\le C\e_0^{|X'|} e^{-\k_0\d'(X')}, \end{equation}
for some $\k_0>0$ and $\e_0$ small enough (here $\d'(X')$ is the rescaled tree length of the coarse
set $X'\subset \L'$, i.e., it is 
the number of nearest neighbor edges of the smallest tree on $\L'$ that covers $X'$).
In the following, we will prove that under the assumption of the Lemma, the polymers' activities 
$K^{(\L)}_q(X)$ satisfy 
\be 
|K_q^{(\L)}(X)|\le \e_1 \e^{|X'|-1}\;,\quad \e_1:=e^{- \frac{c_0}{6}zk^2}\;, \quad
\e_2:=(zk)^{\frac1{32}}\; ,\quad \e:=\max\{\e_1,\e_2 \}\;,
\label{5.6}\ee
where $c_0$ is the same constant as in Eq.\eqref{4.28}. Using the fact that $X$ is D-connected, 
we see that Eq.(\ref{5.6}) implies Eq.(\ref{polym-bound}) with $\e_0=e^{-\k_0}=\e^{1/2}$. Therefore, 
by \cite[Proposition 7.1.1]{GBG}, we get 
\be 
\log\frac{Z(\L|q)}{Z^q(\L)}=\sum_{{\cal X}\subseteq \L}\Big[\prod_{X\in{\cal X}}K_q^{(\L)}(X)\Big]
\phi^T({\cal X})\;.
\label{4.29bis}\ee
Combining this equation with Eq.(\ref{04.3}) gives Eq.(\ref{4.29}).

The rest of this section is devoted to the proof of  Eq.(\ref{5.6}).
The polymer's activity Eq.(\ref{5.5}) can be rewritten as
\bea \label{polym-act}
K_q^{(\L)}(X) &=&   \sum_{\substack{X_{0},X_1 \subseteq X:\\
X_0\cup X_1=X} } \
 \sum_{n\ge 1} \ 
 \sum_{\substack{
\dpr=\{\g_1,\ldots,\g_n\}\in\CC(\L,q):\\
\cup_{i=1}^n\G_i=X_0}}  \Big[     \prod_{j=1}^{n} \z_q(\g_j) \Big] \cdot\\
&&\cdot \sum_{Q\subseteq X_1} \sum_{p\ge 0}\
\sum_{\substack{
\{Y_1,\ldots,Y_p\}:\ Y_i\in\Y^q_{X_0}\\
\cup_i \supp(Y_i)=Q\\ \cup_i\lis Y_i=X_1}}  
 \left [     \prod_{i=1}^{p} ( e^{\FF(Y_j)}-1) \right]\;. \nn
\eea
Note that since all the tiles in $Q$ belong 
to the peel of some contour then $Q\cap X_{0}=\emptyset $. On the other hand, the sets $X_{0}$ and $X_{1}$
{\em may very well overlap}  $X_{0}\cap X_{1}\neq \emptyset $ in general (see Fig.\ref{fig4} for an example). 
Moreover, once $X_{0}$ is fixed, the supports of the contours are automatically fixed too, since they must
be the D-connected components of $X_{0}$. Then we can rewrite the sum as
\begin{align}\label{polym-act1}
K_q^{(\L)}(X) &=\sum_{\emptyset\neq X_{0}\subset X }\
\sum_{\substack{\{\g_1,\ldots,\g_n\}\in\CC(\L,q):\\ {\rm supp} (\gamma_{i} )=\G_i(X_0)}}
  \Big[ \prod_{j=1}^{n} \z_q(\g_j) \Big] \cdot \\
& \quad \cdot  \sum_{\substack{X_1 \subseteq X:\\
X_0\cup X_1=X} } \sum_{Q\subseteq X_1} \sum_{p\ge 0}\
\sum_{\substack{
\{Y_1,\ldots,Y_p\}:\ Y_i\in\Y^q_{X_0}\\
\cup_i \supp(Y_i)=Q\\ \cup_i\lis Y_i=X_1}}  
 \Big[     \prod_{i=1}^{p} ( e^{\FF(Y_j)}-1) \Big] \nonumber
\end{align}
where in the second sum $\G_i(X_0)$ are the maximally D-connected components 
of $X_0$, which must be well D-disconnected (otherwise the corresponding contribution to 
the activity is zero).

Now, note that $\FF(Y)$ given in  Eq.(\ref{5.2}) is at least of order $n$ (with $n\ge 2$) in $z$, by the very 
definition of  the characteristic function $F_\D$. In fact, $F_\D(R)$ is either equal to $f_\D$ or to $f_\D+g_\D(1-f_\D)$;
therefore, using the definitions of $f_\D$ and $g_\D$, Eqs.(\ref{eq:f})-(\ref{eq:g}), we see that 
$F_\D(R)$ is different from zero only if $R$ contains a rod
belonging to $A_{\g(\D)}(\D)$. Now recall that, as already observed after Eq.(\ref{eq:g}), 
distinct tiles $\D_1\neq\D_2$ correspond to distinct sets $A_{\g(\D_1)}(\D_1)$ and 
$A_{\g(\D_2)}(\D_2)$, such that $A_{\g(\D_1)}(\D_1)\cap A_{\g(\D_2)}(\D_2)=\emptyset$
(here $\gamma (\Delta_{i})$ is the contour whose peel $\Delta_{i}$ belongs to, 
$\Delta_{i}\in P_{\Gamma_{i}}$: since the peels of different contours are disconnected, the contour $
\gamma (\Delta_{i}) $ is unique). Therefore, 
the r.h.s. of Eq.(\ref{5.2}) is non zero only if $R$ contains at least $n$ distinct rods. 
Using Eq.(\ref{R4.9}), we find that, if $Y=\{\D_{\x_1},\ldots,\D_{\x_m}\}$ with $\D_{\x_1}<\cdots<\D_{\x_n}$,
\bea
|\FF(Y)| \leq 
\sum_{\substack{R\in \O^q_{\L}:\ |R|\ge 2\\
V (R)\cap  \D_1\neq \emptyset,\\
 {\rm diam}({\rm supp}(R) )\ge \diam(Y)}}|z|^{|R|}|\f^T(R)|&\le&  2Cz\ell^2(Czk)^{\max\{\frac{\diam(Y)}{k-1}-1,1\}}\nn\\ &\le & C' zk^2 (zk)^{\a\cdot \diam'(Y)} \;,
\label{5.3}
\eea
where 
$\diam'(Y)=|\x_n-\x_1|/\ell$ is the rescaled diameter of the set $\cup_{\D\in Y}\D$, and $\a$ can be 
chosen to be $\a=1/4$.
Using this bound and  the fact that $|e^x-1|\le |x|e^{|x|}$, we find:
\bea &&\label{bound1}
\sum_{\substack{\{Y_1,\ldots,Y_p\}:\ Y_i\in\Y^q_{X_0}\\ 
\cup_i \supp(Y_i)=Q\\ \cup_i\lis Y_i=X_1}} 
    \prod_{i=1}^{p}  \left | e^{\FF(Y_j)}-1 \right | \le\\
    &&\le 
\sum_{\substack{\{Y_1,\ldots,Y_p\}:\ Y_i\in\Y^q_{X_0}\\ 
\cup_i \supp(Y_i)=Q\\ \cup_i\lis Y_i=X_1}}    \prod_{j=1}^p \left  \{  C' zk^2(zk)^{\alpha \cdot\diam'(Y_j)}
e^{ \left [ 
C' zk^2(zk)^{\alpha \cdot\diam'(Y_j)}\right]  }  \right \}\nn\eea
Now, since the choice of $Y$ only  depends on the union of the contours' supports $X_0$, in 
Eq.(\ref{polym-act1}) we can start with performing the sums
over the contours' spin attributions, rod configurations and internal colors.
Using the bound Eq.\eqref{4.28} on the contours activities, we get
\begin{align}\label{actbound1}
& \sum_{\substack{\{\g_1,\ldots,\g_n\}\in\CC(\L,q):\\ {\rm supp} (\gamma_{i} )=\G_i(X_0)}}
  \Big[ \prod_{j=1}^{n} \z_q(\g_j) \Big]\le   \prod_{j=1}^{n}  \Big[    
 \sum_{\substack{\gamma_{j}:\\   \supp (\gamma_{j} )=\Gamma_{j}(X_0)}}  |\z_q(\g_j)|\Big] \\
&\qquad \qquad   \leq    \hspace{-0.1cm} 
  \prod_{j=1}^{n}  6^{|\Gamma'_{j}|} e^{-c_0\cdot zk^2|\Gamma_{j}'|} =
     6^{|X_{0}'|}  e^{-c_0\cdot zk^2 |X_{0}'|} \nonumber
\end{align}
where the factor $6^{|\Gamma'|}$ bounds the sums over $\s_\g$ and $\ul m_{int}$ at $\G$ fixed. Putting
these results together into Eq.(\ref{polym-act1}), we find
 \bea  
|K_q^{(\L)}(X)|
&\le& \sum_{\emptyset\neq X_{0}\subset X }\   6^{|X_{0}'|}  e^{-c_0\cdot zk^2 |X_{0}'|} \sum_{\substack{X_1 \subseteq X:\\
X_0\cup X_1=X} } \sum_{Q\subseteq X_1} \sum_{p\ge 0}\
\sum_{\substack{
\{Y_1,\ldots,Y_p\}:\ Y_i\in\Y^q_{X_0}\\
\cup_i \supp(Y_i)=Q\\ \cup_i\lis Y_i=X_1}}  \cdot\nonumber
\\
&&\cdot\prod_{j=1}^p\left \{  C' zk^2(zk)^{\alpha \cdot\diam'(Y_j)}e^{\left[ C' zk^2(zk)^{\alpha \cdot\diam'(Y_j)}  \right]   }
 \right\} \;,
\label{5.7}\eea

Now, note that: (i) $\sum_{j=1}^p \diam'(Y_j)\ge |X_1'|-1\ge|X_1'|/2$; (ii) $|X_0'|\ge |Q'|$, because every tile 
in $Q=\cup_i \supp(Y_i)$ belongs to the peel of $X_0$;
\be
{\rm (iii)}\quad \sum_{j=1}^p(zk)^{\alpha \cdot\diam'(Y_j)}\le \sum_{\x\in Q'}\sum_{\lis Y'\ni \x}(zk)^{\alpha |\lis Y'|}
\le C''|Q'|(zk)^{\alpha } \;\nn
\ee
Plugging these estimates into Eq.(\ref{5.7}) we find
\bea  &&
|K_q^{(\L)}(X)|\le
\sum_{\emptyset\neq X_0\subseteq X}
 (6e^{-\frac{c_0}2\cdot zk^2})^{|X_0'|}  \sum_{\substack{X_1\subseteq X:\\
X_0\cup X_1=X}} (zk)^{\frac{\alpha }4|X_1'|}\cdot\label{5.8}\\
&&\cdot\sum_{Q\subseteq X_1} 
e^{-zk^2|Q'|(\frac{c_0}2-C'C''(zk)^{\alpha })}
\sum_{p\ge 0}\sum_{\substack{\{Y_1,\ldots,Y_p\}:\ Y_i\in\Y^q_{X_0}\\
\cup_i \supp(Y_i)=Q\\ \cup_i\lis Y_i=X_1}}
\prod_{j=1}^p\Big[ C' zk^2(zk)^{\frac{\alpha }2\diam'(Y_j)}\Big]\;,
\nn\eea
which can be further bounded by:
\bea  |K_q^{(\L)}(X)|&\le& 
\sum_{\emptyset\neq X_0\subseteq X}
 e^{-\frac{c_0}3zk^2|X_0'|}  \sum_{\substack{X_1\subseteq X:\\
X_0\cup X_1=X}}  (zk)^{\frac{\alpha }4|X_1'|}\cdot\label{5.9}\\
&&\cdot\sum_{Q\subseteq X_1} 
e^{-\frac{c_0}3zk^2|Q'|}
\sum_{p\ge 0}\frac1{p!}\Big[ C' zk^2
\sum_{\substack{A\cap Q\neq\emptyset\\  |A'|\geq 2}}(zk)^{\frac{\alpha }2\d'(A')}\Big]^p\;,\nn\eea
where in the last sum $A$ is a generic subset of $\L$ consisting of a union of tiles, and $\d'(A')$ is its 
rescaled tree length. The expression in square brackets in the second line is bounded above by 
$C'' zk^2|Q'|(zk)^{\frac{\alpha }2}$, so that 
\bea  
|K_q^{(\L)}(X)|&\le&
\sum_{\emptyset\neq X_0\subseteq X}
e^{-\frac{c_0}3zk^2|X_0'|}  \sum_{\substack{X_1\subseteq X:\\
X_0\cup X_1=X}}
(zk)^{\frac{\alpha }4|X_1'|}\sum_{Q\subseteq X_1} 
e^{-zk^2|Q'| (\frac{c_0}3-C''(zk)^{\frac{\alpha }{2}})}\nn\\
&\le&
\sum_{\emptyset\neq X_0\subseteq X}
e^{-\frac{c_0}3zk^2|X_0'|} \sum_{\substack{X_1\subseteq X:\\
X_0\cup X_1=X}} (zk)^{\frac{\alpha }4|X_1'|}\sum_{Q\subseteq X_1} 
e^{-\frac{c_0}4zk^2|Q'|}\;.
\label{5.10}\eea
The last sum can be rewritten as $\sum_{Q\subseteq X_1} 
e^{-\frac{c_0}4zk^2|Q'|}= (1+e^{-\frac{c_0}4zk^2})^{|X_1'|}$, so that, defining 
$\tilde\e_1:=e^{-\frac{c_0}3zk^2}$, $\tilde\e_2:=(zk)^{\frac{\alpha }4}$ and $\tilde\e:=\max\{\tilde\e_1,
\tilde\e_2\}$:
\bea
|K_q^{(\L)}(X)| &\leq&
\sum_{\emptyset\neq X_0\subseteq X}\tilde\e_1^{|X_0'|}  \sum_{\substack{X_1\subseteq X:\\
X_0\cup X_1=X}}\left ( (1+\tilde\e_{1}^{\frac{3}{4}})\tilde\e_2\right )^{|X_{1}'|} \nn\\
&\le& \sum_{\emptyset\neq X_0\subseteq X}\tilde\e_1^{|X_0'|}  \sum_{\substack{X_1\subseteq X:\\
X_0\cup X_1=X}}\left (2\tilde\e_2\right )^{|X_{1}'|}= \left(\tilde\e_1+ 2\tilde\e_1 \tilde\e_2   +  
2\tilde\e_2  \right)^{|X'|}- (2\tilde\e_2)^{|X'|}
\nn\\
& \leq& |X'| \tilde\e_{1}(1+ 2\tilde\e_2)  
\left(\tilde\e_1+ \tilde\e_1 2\tilde\e_2 + 2\tilde\e_2\right)^{|X'|-1} \leq \tilde\e_1 (\sqrt{\tilde\e})^{|X'|-1}
\label{5.11}\eea
where $\tilde\e= {\rm max } \{\tilde\e_1,\tilde\e_2\}$. Setting $\e_{1}=  \sqrt{\tilde\e_1}$,  
$\e_{2}=  \sqrt{\tilde\e_2}$ and recalling that $\a=\frac14$, 
we obtain the desired estimate on $K^{(\Lambda )}_{q}(X)$. This concludes the proof of the lemma.\qed

{\bf Remark.} The dependence of the activities $K_q^{(\L)}(X)$ on $\L$ is
inherited from the 
constraint that $X$ must be separated from $\L^c$ by at least one smoothing square, and by the fact 
that the quantities $\z(\g)$ and $\FF(Y)$ themselves are $\L$-dependent, simply because their 
definitions involve sums over rods collections in $\O^q_\L$. However, this dependence is very weak: 
in fact, if $K_q(X)$ is the infinite 
volume limit of $K_q^{(\L)}(X)$, we have:
\be \big|K_q^{(\L)}(X)-K_q(X)\big|\le\big( \e_1\e^{|X'|-1}\big)^{1/2}\e^{c'\cdot\dist'(X',\L^{'}_c)}\;,\label{5.12}\ee
for some $c'>0$. The proof of Eq.(\ref{5.12}) proceeds along the same lines used to prove Eq.(\ref{5.6}) 
and, therefore, we will not belabor the details of this computation. 


\section{The activity of the contours}\label{sec6}
\setcounter{equation}{0}
\renewcommand{\theequation}{\ref{sec6}.\arabic{equation}} 


In this section we prove the assumption Eq.(\ref{4.28}) used in the proof of Lemma 
\ref{lemma:convergence}. Let us first remind, for the reader's convenience, 
the definition of $\z_q(\g)$:
\begin{equation}
\zeta_{q} (\gamma) =  \z^0_q(\g)\,\exp\Big\{-\sum_{R\in \O^q_\L}\f^T(R)z^{|R|}
\sum_{\D\subseteq P_\G}F_\D(R)\Big\}\;,\label{6.1}
\end{equation} 
where 
\be \z^0_q(\g)=\frac{\bar\varphi (R_{\gamma })  }{Z^{q}(\Gamma ) } \prod_{j=1}^{h_\G}  
\frac{ Z_{\gamma } (\Int_{j}\G |m^j_{int})}{ Z (\Int_j\G |q )}\;.\label{6.2}\ee
By using the same considerations used to get the bound Eq.(\ref{5.3}), we see that the 
expression in braces in the r.h.s. of Eq.(\ref{6.1}) is equal to a contribution of order one in $z$ plus a rest, 
which is bounded in absolute value by 
$Czk^2|\G'|(zk)^\alpha $. On the other hand, the contribution of order one in $z$ is equal to 
$-z\sum_{R\in\O^q_\L:\ |R|=1} \sum_{\D\subseteq P_\G}F_\D(R)$,
 which is {\it negative}, simply because $F_\D\ge 0$. Therefore,
\be 
|\zeta_{q} (\gamma)| \le  |\z^0_q(\g)| e^{Czk^2|\G'|(zk)^{\alpha }}\;,
\label{6.3}\ee
which makes apparent that, in order to prove Eq.(\ref{4.28}), we need to prove an 
analogous bound for $\z^0_q(\g)$. By definition,  $Z_\g(X|m)\le Z(X|m)$, so that 
\be  
|\z^0_q(\g)|\le|{\lis\z}^0_q(\g)|\prod_{j=1}^{h_\G}\max\Big\{1,\frac{Z(\Int_j\G|-q)}{Z(\Int_j\G|q)}\Big\}\;,\qquad 
\lis\z^0_q(\g):=\frac{\bar\varphi (R_{\gamma })  }{Z^{q}(\Gamma ) }\;.
\label{6.4}\ee
The estimate that we need on the quantities $\lis\z^0_q(\g)$ and $\frac{Z(\Int_j\G|-q)}{Z(\Int_j\G|q)}$
is summarized in the following two lemmas.

\begin{lemma}\label{lemma:activities}
Let $zk$ and $(zk^2)^{-1}$ be small enough. Then
\begin{equation}\label{6.5}
\sup_{\s_\g}\!^*\sum_{R_\g\in\O_\G(\s_\g)}|\lis\z^0_q(\g)|\le e^{-2c_{0}\, zk^2|\G'|}\;,
\end{equation}
where the $*$ on the sup reminds the constraint that all the smoothing squares in $\G$
must have a non-zero intersection with at least one bad sampling square, and $c_0=5*10^{-4}$.\end{lemma}

\noindent{\bf Remark.} The specific choice of $c_0$ in the lemma comes from Eq.(\ref{6.9}) below. 
It is related to the size of the smoothing squares, to the number of  zero spins and to the number of pairs 
of neighboring spins with opposite sign that can appear in a contour (as explained below, 
it comes from the remark that every smoothing square - which contains 64 tiles - in a contour 
must intersect at least one bad sampling square - of size $\ell^2\ge k^2/4$). 

\begin{lemma}\label{lemma:ratios}
Let $zk$ and $(zk^2)^{-1}$ be small enough. Then there exist two positive constants $C,c_{1}>0$ such that,
for any simply connected region $X\subset\ZZZ^2$ consisting of a union of smoothing squares,
\begin{equation}\label{6.6} 
e^{-|P_X'|(Czk^2(zk)+\e^{c_{1}})}\le \frac{Z(X|+)}{Z(X|-)}\le e^{|P_X'|(Czk^2(zk)+\e^{c_{1}})}\;,\end{equation}
where $\e$ was defined in Eq.(\ref{5.6}) and $P_X$ is the 1-tile-thick peel of $X$.
\end{lemma}

These two estimates combined with Eq.(\ref{6.3}) give
\begin{align}
\sup_{\s_\g}\!^*\sum_{R_\g\in\O_\G(\s_\g)}|\z_q(\g)| & \le \    e^{-2c_{0}\, zk^2|\G'|} e^{C'zk^2|\G'|(zk)^{\alpha }}\ 
 \prod_{j=1}^{h_\G}e^{|P'_{\Int_j\G}|(Czk^2(zk)+\e^{c_{1}}) }   \nn\\
&
\leq  \  e^{|\Gamma'|(Czk^2(zk)+\e^{c_{1}}) }
e^{C'zk^2|\G'|(zk)^{\alpha }} e^{-2c_{0}\, zk^2|\G'|}  \nn\\
& =\  e^{- zk^2|\G'| \left( 2
c_{0}- C' (zk)^{\alpha } - C (zk)- \frac{\e^{c_{1}}}{ zk^2} \right)}\leq  \  e^{-c_{0} zk^2|\G'|}
\end{align}
 under the only assumptions 
that $zk$ and $(zk^2)^{-1}$ are small enough.
Therefore, these two lemmas imply the convergence
of the cluster expansion Eq.(\ref{4.29}), which completes the computation of the partition function 
of our hard rod system with $q$ boundary conditions. A computation of the correlation functions
based on a similar expansion will be discussed in the next section. The rest of this section is devoted to the 
proofs of Lemma \ref{lemma:activities} and \ref{lemma:ratios}.\\

{\bf Proof of Lemma \ref{lemma:activities}}. Let $\s_\g$ be a spin configuration compatible with the 
fact that $\g$ is a contour. In particular, let us recall that every smoothing square contained in $\G$ has
a non zero intersection with at least one bad sampling square; moreover, by its very definition, each
such bad square must contain either one tile with magnetization equal to $0$, or one pair of
neighboring tiles with magnetizations $+$ and $-$, respectively. Therefore, given $\s_\g$, it is 
possible to exhibit a partition $\PPP$ of $\G$ such that: (i) all the elements of the partition consist either 
of a single tile or of a pair of neighboring tiles with opposite magnetizations $+$ and $-$ (we shall call
such pairs ``domino tiles"); (ii) if $\NN_0$ is the number of single tiles in $\PPP$ with magnetization 
equal to 0 and $\NN_{d}$ is the number of domino tiles in $\PPP$, then $\NN_0+\NN_{d}\ge |\G'|/64$.
The factor 64 comes from the consideration that in $\G$, by definition, we have at least one bad square 
every four smoothing squares, and by the fact that four smoothing squares  contain 64 tiles. 

By the definition of $\bar\f(R_\g)$, we have: $\bar\f(R_\g)\le \prod_{P\in\PPP}\bar\f(R_P)$. 
Moreover, using the standard cluster expansion described after Eq.(\ref{4.4}), we find that 
$Z^q(\G)\ge \prod_{P\in\PPP}Z^q(P)e^{-Czk^2(zk)|\G'|}$. By combining these two bounds we get
\be \sum_{R_\g\in\O_\G(\s_\g)}|\lis\z^0_q(\g)|\le e^{Czk^2(zk)|\G'|}\prod_{P\in\PPP}\Big|\sum_{R_P}\frac{\bar\f(R_P)}{Z^q(P)}\Big|\;,\label{6.7}\ee
where the sum over $R_P$ runs over rods configurations in $\O_P(\cup_{\x\in P'}\s_\x)$. 
Now, if $P$ is a single tile with magnetization either $+$ or $-$,  
then $\sum_{R_P}\frac{\bar\f(R_P)}{Z^q(P)}=1$. 
Moreover, if $P$ is a single tile with magnetization equal to 0, 
then $\sum_{R_P}\frac{\bar\f(R_P)}{Z^q(P)}=-\frac{1}{Z^q(P)}=-e^{-z\ell^2(1+O(zk))}$.

Finally, let us consider the case that $P$ is a domino tile. We assume without loss of generality that
$P=\{\D_{\x_1},\D_{\x_2}\}$, with $\x_2-\x_1=(\ell,0)$, and $\s_{\x_1}=-\s_{\x_2}=+$.
Since the rods interact via a hard core, $\bar\f(R_{\x_1},R_{\x_2})$ is different from zero only if at least 
one of the two rod configurations $R_{\x_1}$ and $R_{\x_2}$ is {\it untypical}: here we say that 
$R_{\x_1}$ is untypical if it does not contain any  rod in the right half of $\D_{\x_1}$ and, similarly, that
$R_{\x_2}$ is untypical if it does not contain any rod in the left half of $\D_{\x_2}$. Therefore,
\be \sum_{R_P}\frac{\bar\f(R_P)}{Z^q(P)}\le e^{Czk^2(zk)}\Big[\sum_{\substack{
R_{\x_1}\in\O_{\D_{\x_1}}^+:\\ R_{\x_1}\ {\rm untypical}}}
\frac{\bar\f(R_{\x_1})}{Z^+(\D_{\x_1})}+
\sum_{\substack{
R_{\x_2}\in\O_{\D_{\x_2}}^-:\\ R_{\x_2}\ {\rm untypical}}}
\frac{\bar\f(R_{\x_2})}{Z^-(\D_{\x_2})}\Big]\;,\label{6.8}\ee
where we used that $\sum_{R\in\O^q_\D}\bar\f(R)=Z^q(\D)$. Eq.(\ref{6.8}) can be rewritten and 
estimated (defining $\D^L_{\x_1}$ to be the left half of $\D_{\x_1}$) as
\be \sum_{R_P}\frac{\bar\f(R_P)}{Z^q(P)}\le 2 e^{Czk^2(zk)}\frac{Z^+(\D^L_{\x_1})}{Z^+(\D_{\x_1})}\le
2 e^{C'zk^2(zk)} e^{-z\ell^2/2}\;.\label{6.9}\ee

Plugging the bounds on $\sum_{R_P}\frac{\bar\f(R_P)}{Z^q(P)}$ into Eq.(\ref{6.7}) gives:
\bea \sum_{R_\g\in\O_\G(\s_\g)}|\lis\z^0_q(\g)|&\le& e^{Czk^2(zk)|\G'|}
e^{-z\ell^2(1-Czk)(\NN_0+\frac12\NN_d)}\nn\\
&\le& e^{-z\ell^2(1-C'zk)|\G'|/128}\;,\label{6.10}\eea
where in the last line we used the bound $\NN_0+\NN_d\ge |\G'|/64$. 
Using $\ell\geq k/2$ we obtain Eq.(\ref{6.5}) so the proof of the lemma is complete. \qed
\vskip.2truecm

{\bf Proof of Lemma \ref{lemma:ratios}.} We proceed by induction on the size of $X$. If $X$ is so small
that it cannot contain contours D-disconnected from $X^c$, then 
\be \frac{Z(X|+)}{Z(X|-)}= \frac{Z^+(X)}{Z^-(X)}=\exp\Big\{\sum_{R\in\O^+_X}\f^T(R)z^{|R|}-
\sum_{R\in\O^-_X}\f^T(R)z^{|R|}\Big\}\;.\label{6.11}\ee
Let $V(R)$ be the union of the centers of the rods in $R$ and let $R\in\O^q_X$. Since the orientation of all rods in 
$R$ is fixed,  $V(R)$ identifies uniquely the rod configuration. Then
\bea 
\sum_{R\in\O^q_X}\f^T(R)z^{|R|}&=& \sum_{R\in\O^q_X}  \sum_{x\in V (R)}\frac{\f^T(R)z^{|R|}}{|V(R)|} =
  \sum_{x\in X}\sum_{\substack{R\in\O^q_X\\
V(R)\ni x}}\frac{\f^T(R)z^{|R|}}{|V(R)|}\label{6.11a}\\
&=& \sum_{x\in X}\sum_{\substack{R\in\O^q_{\mathbb{Z}^2}\\
V(R)\ni x}}\frac{\f^T(R)z^{|R|}}{|V(R)|}-\sum_{x\in X}\sum_{\substack{R\in\O^q_{\mathbb{Z}^2}\setminus\O^q_X\\
V(R)\ni x}}\frac{\f^T(R)z^{|R|}}{|V(R)|}\;.
\nn\eea
The first sum in the second line is equal to 
\be \sum_{x\in X}\sum_{\substack{R\in\O^q_{\mathbb{Z}^2}\\
V(R)\ni x}}\frac{\f^T(R)z^{|R|}}{|V(R)|}=|X|s(z)\;,\label{6.11b}\ee
where 
\be s(z):=\sum_{\substack{R\in\O^q_{\mathbb{Z}^2}\\
V(R)\ni x}}\frac{\f^T(R)z^{|R|}}{|V(R)|}\label{6.12}\ee
is an analytic function of $z$, of the form $s(z)=z(1+O(zk))$, {\em independent of $q$ and $x$}. The second 
sum in the second line of Eq.(\ref{6.11a}) involves rod configurations containing at least one rod 
belonging to $X$ and one belonging to $X^c$. Therefore, 
it is of order at least 2 in $z$ and scales like the boundary of $X$:
\be
\Big| \sum_{x\in X}\sum_{\substack{R\in\O^q_{\mathbb{Z}^2}\setminus\O^q_X\\
V(R)\ni x}}\frac{\f^T(R)z^{|R|}}{|V(R)|}\Big|\le C_1zk^2(zk)|P_X'|\;,
\label{6.13} \ee
for a suitable constant $C_1>0$, independent of $q$. Plugging Eqs.(\ref{6.11a})--(\ref{6.13})
into Eq.(\ref{6.11}) gives:
\be  
\frac{Z(X|+)}{Z(X|-)}=\exp\Big\{-\sum_{x\in X}\sum_{\substack{R\in\O^+_{\mathbb{Z}^2}\setminus\O^+_X\\
V(R)\ni x}}\frac{\f^T(R)z^{|R|}}{|V(R)|}+\sum_{x\in X}\sum_{\substack{R\in\O^-_{\mathbb{Z}^2}\setminus\O^-_X\\
V(R)\ni x}}\frac{\f^T(R)z^{|R|}}{|V(R)|}\Big\}\;,\label{6.14}\ee 
which is bounded from above and below by $e^{2C_1zk^2(zk)|P_X'|}$ and $e^{-2C_1zk^2(zk)|P_X'|}$,
respectively. Setting $C\ge 2C_1$,  this proves the inductive hypothesis Eq.(\ref{6.6}) at the first step, 
i.e., for regions $X$ small enough.

Let us now assume the validity of Eq.(\ref{6.6}) for all the
regions of size strictly smaller than $\L_0$, and let us prove it for $\L_0$. As explained in Section
\ref{sec5}, $Z(\L_0|q)$ admits the cluster expansion Eq.(\ref{4.29}) involving polymers $X$
that are D-disconnected from $\L_0^c$, whose activities are defined in Eq.(\ref{5.5}). In particular,
the cluster expansion is convergent provided that $\z_q(\g)$ is bounded as in Eq.(\ref{4.28}).
Now, note that the interiors of the contours $\g_i$ involved in the cluster expansion for $Z(X_0|q)$
via Eqs.(\ref{4.29}) and (\ref{5.5})
have all sizes strictly smaller than $\L_0$. Therefore, using the inductive hypothesis, 
the product $\max\Big\{1,\frac{Z(\Int_j\G|-q)}{Z(\Int_j\G|q)}\Big\}$ in Eq.(\ref{6.4}) can be bounded from
 above by $e^{|\G'|(Czk^2(zk)+\e^c)}$ that, if combined with Eqs.(\ref{6.3}), (\ref{6.5}), implies 
 Eq.(\ref{4.28}) for all the the contours $\g_i$ involved in the cluster expansion for 
 $Z(X_0|q)$. We can then write:
\be 
\frac{Z(\L_0|+)}{Z(\L_0|-)}=  \frac{Z^+(\L_0)}{Z^-(\L_0)} \exp\Big\{
\sum_{{\cal X}\subseteq \L_0}\Big[K^{(\L_0)}_+({\cal X})-K^{(\L_0)}_-({\cal X})\Big]
\phi^T({\cal X})\Big\}\;,\label{6.15}\ee
where $K^{(\L_0)}_q({\cal X})=\prod_{X\in{\cal X}}K^{(\L_0)}_q({X})$ and 
 $K^{(\L_0)}_q(X)$ admits the bound Eq.(\ref{5.6}). The first factor in the r.h.s. of Eq.(\ref{6.15})
 is rewritten as in Eq.(\ref{6.14}) and is bounded from above and below by $e^{2C_1zk^2(zk)|P_{\L_0}'|}$ and $e^{-2C_1zk^2(zk)|P_{\L_0}'|}$,
respectively, exactly in the same way as Eq.(\ref{6.15}) itself. 

The second factor in the r.h.s. of Eq.(\ref{6.15}) can be bounded as follows. 
 We rewrite 
 \bea &&  \exp\Big\{
\sum_{{\cal X}\subseteq \L_0}\Big[K^{(\L_0)}_+({\cal X})-K^{(\L_0)}_-({\cal X})\Big]
\phi^T({\cal X})\Big\}= \label{6.17}\\
&& =\exp\Big\{\sum_{\substack{{\cal X}\subseteq \L_0\\ q=\pm}}qK_q({\cal X})
\phi^T({\cal X})\Big\}\cdot \exp\Big\{
\sum_{\substack{{\cal X}\subseteq \L_0\\ q=\pm}}q\Big[K^{(\L_0)}_q({\cal X})-K_q({\cal X})\Big]
\phi^T({\cal X})\Big\}\;,\nn\eea
where $K_q({\cal X})=\prod_{X\in{\cal X}}K_q({X})$. Now 
\begin{equation}
\prod_{j=1}^{n} K_q^{(\L_0)}(X_{j}) - \prod_{j=1}^{n} K_q(X_{j})  = \sum_{m=1}^{n}  
\prod_{j=1}^{m-1} K_q(X_{j}) \ [ K_q^{(\L_0)}(X_{m})- K_q(X_{m})   ] \prod_{j=m+1}^{n} K^{(\L_0)}_q(X_{j})
\end{equation}
then  using Eq.(\ref{5.12}) and \eqref{5.6} , we have
\begin{align}
|K_q^{(\L_0)}({\cal X})-K_q({\cal X})|&\le 
\sum_{m=1}^{n}    (\sqrt{\e})^{|X'_m|}  \e^{c'\dist'(X'_{m},\L_{0,c}')}   \prod_{j\neq m} \e^{|X'_j|}
\nn\\
&  \leq 
\e^{c'\dist'({\cal X},\L_{0,c}')}\prod_{X\in{\cal X}}\e^{\frac{|X'|}{2}}\;,
 \label{6.16}
\end{align}
Therefore, the second factor in the second line of Eq.(\ref{6.17})
 can be bounded from above and below by $e^{\e^{c_2}|P_{\L_0}'|}$ and
 $e^{-\e^{c_2}|P_{\L_0}'|}$,
respectively for a suitable constant $c_{2}$. 
We are left with the first factor in the second line of Eq.(\ref{6.17}), which involves the 
partition sum 
\be 
\sum_{{\cal X}\subseteq \L_0}K_q({\cal X})
\phi^T({\cal X})=\sum_{\x\in \L_0'}\sum_{\substack{{\cal X}\supseteq \D_\x\\ {\cal X}\subseteq\L_0}}
\frac{K_q({\cal X})\phi^T({\cal X})}{|{\cal X}'|}\;,\label{6.18}\ee
where $|{\cal X}'|$ is number of tiles in $\cup_{X\in{\cal X}}X$. Eq.(\ref{6.18}) can be further rewritten as
\be  \sum_{{\cal X}\subseteq \L_0}K_q({\cal X})
\phi^T({\cal X})=|\L_0'|{\cal S} +\sum_{\x\in \L_0'}\sum_{\substack{{\cal X}\supseteq \D_\x\\ {\cal X}\cap\L_0^c\neq\emptyset}}
\frac{K_q({\cal X})\phi^T({\cal X})}{|{\cal X}'|}\;,\label{6.19}\ee
where 
\be {\cal S}:=\sum_{\substack{{\cal X}\supseteq \D_\x\\ {\cal X}\subseteq{\mathbb Z}^2}}
\frac{K_q({\cal X})\phi^T({\cal X})}{|{\cal X}'|}\;,\label{6.20}\ee
is independent of $q$ and $\x$. The second term in the r.h.s. of Eq.(\ref{6.19})
is bounded in absolute value from above by $|P_{\L_0}'|\e^{c_3}$ for a suitable $c_3>0$; 
therefore,
\be \exp\Big\{\sum_{\substack{{\cal X}\subseteq \L_0\\ q=\pm}}qK_q({\cal X})
\phi^T({\cal X})\Big\}=\exp\Big\{
\sum_{\substack{\x\in \L_0'\\ q=\pm}}\sum_{\substack{{\cal X}\supseteq \D_\x\\ {\cal X}\cap\L_0^c\neq\emptyset}}q
\frac{K_q({\cal X})\phi^T({\cal X})}{|{\cal X}'|}\Big\}\le e^{2|P_{\L_0'}|\e^{c_3}}\label{6.21}\ee
and is bounded from below by $e^{-2|P_{\L_0'}|\e^{c_3}}$.
Choosing  $c_1$ such that $\e^{c_1}\ge \e^{c_2}+2\e^{c_3}$ this completes the inductive proof of Eq.(\ref{6.6}).\qed


\section{Existence of nematic order}\label{sec7}
\setcounter{equation}{0}
\renewcommand{\theequation}{\ref{sec7}.\arabic{equation}} 

 
In this section we prove Theorem \ref{thm1}. We start by proving Eq.(\ref{3.9}). The probability that 
the tile centered at $\x_0$ has magnetization $-q$ in the presence of boundary conditions $q$ can be
written as
\be \media{\c_{\x_0}^{-q}}_\L^q=\frac{\dpr}{\dpr z_0}\log Z_{z_0}(\L|q)\Big|_{z_0=1}\;,\label{7.1}\ee
where $Z_{z_0}(\L|q)$ is defined in a way completely analogous to Eqs.(\ref{3.4})-(\ref{3.6}), with the only difference that the activity $\z(\x)$ in Eq.(\ref{3.4}) is replaced by $\tilde\z(\x)$, where $\tilde\z(\x)=
\z(\x)$ if $\x\neq\x_0$, while 
\be \tilde\z(\x_0)= \left\{
\begin{array}{ll}
z^{|R_{\xi }|} & \mbox{if}\  \sigma_{\xi }=q\\
z_0z^{|R_{\xi }|} & \mbox{if}\  \sigma_{\xi }=-q\\
-1  & \mbox{if}\  \sigma_{\xi }=0.\\
\end{array}\right.\ee
The change of $\z(\x)$ into $\tilde \z(\x)$ induces a corresponding change of $\z_q(\g)$ and 
$K^{(\L)}_q(X)$ 
into $\tilde\z_q(\g)$ and $\tilde K^{(\L)}_q(X)$, respectively. The activity $\tilde K^{(\L)}_q(X)$ admits the 
same bound Eq.(\ref{5.6}) (possibly with a slightly different constant $c''$), uniformly in $z_0$ for $z_0$ close to $1$, and it depends explicitly on 
$z_0$ only if $X\supseteq \D_{\x_0}$.
In such a case, the derivative of $\tilde K^{(\L)}_q(X)$ with respect to $z_0$ is bounded by 
$\sqrt{\e_1\e^{|X'|-1}}$, uniformly in $z_0$ for $z_0$ close to $1$. 

The logarithm of the 
modified partition function admits a convergent cluster expansion analogous to Eq.(\ref{4.29}):
\be \log \frac{Z_{z_0}(\L|q)}{Z^q(\L)}=
\sum_{{\cal X}\subseteq \L}\tilde K^{(\L)}_q({\cal X})\phi^T({\cal X})\;,
\label{7.1a}\ee
so that 
\be \media{\c_{\x_0}^{-q}}_\L^q=
\sum_{{\cal X}\subseteq \L}\dpr_{z_0}\tilde K^{(\L)}_q({\cal X})\phi^T({\cal X})\Big|_{z_0=1}\;.\label{7.2}\ee
The sum in the r.h.s. of Eq.(\ref{7.2}) is exponentially convergent for $\e$ small enough, and it only
involves polymer configurations containing $\D_{\x_0}$, simply because $\tilde K^{(\L)}_q(X)$ is independent of $z_0$ whenever $\D_{\x_0}\cap X=\emptyset$. Therefore, 
\bea \media{\c_{\x_0}^{-q}}_\L^q\le 
\sum_{{\cal X}\subseteq \L}|\phi^T({\cal X})|\cdot|\dpr_{z_0}\tilde K^{(\L)}_q({\cal X})|_{z_0=1}&\le&
(\e_1)^{\frac14}\sum_{{\cal X}\supseteq \D_{\x_0}}|\phi^T({\cal X})|\prod_{X\in{\cal X}}\e^{\frac14|X'|}\nn\\
&\le& (\const.)
(\e_1)^{\frac14}\;,\label{7.3}\eea
which proves Eq.(\ref{3.9}).

In order to compute the density-density correlation functions we proceed in a similar fashion.
We replace the activity $z$ of a rod $r$ centered at $x$ by $z_x$ and we define $\tilde Z_{\bf z}(\L|q)$
to be the modified partition function with boundary conditions $q$ and variable rod activities 
${\bf z}=\{z_x\}_{x\in\L}$. Correspondingly, we rewrite:
\bea && \media{n_x}^q_\L=z\dpr_{z_x}\log \tilde Z_{\bf z}(\L|q)\Big|_{{\bf z}= z}\;,\nn\\
&& 
\media{n_x n_y}^q_\L-\media{n_x}^q_\L\media{n_y}^q_\L=z^2
\dpr_{z_x}\dpr_{z_y}\log \tilde Z_{\bf z}(\L|q)\Big|_{{\bf z}= z}\;,\label{7.4}\eea
where ${\bf z}=z$ means that $z_x=z$, $\forall x\in\L$; the higher order density correlation functions 
have  a similar representation. Once again, $\log \tilde Z_{\bf z}(\L|q)$ 
admits a cluster expansion completely analogous to $\log Z(\L|q)$:
\be \log \tilde Z_{\bf z}(\L|q)=\sum_{R\in \Omega_{\L}^{q}}\big[\prod_{r\in R}z_{x(r)}\big]{\varphi }^T (R)+
\sum_{{\cal X}\subseteq \L}\tilde K^{(\L)}_{q,{\bf z}}({\cal X})\phi^T({\cal X})\Big|_{{\bf z}=z}\;,\label{7.5}\ee
where $x(r)$ is the center of $r$. Moreover, $\tilde K^{(\L)}_{q,{\bf z}}(X)$, together with its derivatives 
with respect to $z_x$ and/or $z_y$,  admit the 
same bound Eq.(\ref{5.6}), possibly with a different constant $c''$; the derivative of 
$\tilde K^{(\L)}_{q,{\bf z}}(X)$
with respect to $z_x$ and/or $z_y$ is different from zero only if $X\ni x$ and/or $X\ni y$.
Therefore,
\bea && \media{n_x}^q_\L=\sum_{R\in \Omega_{\L}^{q}}z^{|R|}R(r(x))
{\varphi }^T (R)+
\sum_{{\cal X}\subseteq \L}\dpr_{z_x}\tilde K^{(\L)}_{q,{\bf z}}({\cal X})\phi^T({\cal X})\Big|_{{\bf z}=z}
\;,\nn\\
&& \media{n_xn_y}^q_\L-\media{n_x}^q_\L\media{n_y}^q_\L=\label{7.6}\\
&&\qquad =
\sum_{R\in \Omega_{\L}^{q}}z^{|R|}R(r(x))R(r(y)){\varphi }^T (R)+
\sum_{{\cal X}\subseteq \L}\dpr^2_{z_x\,z_y}\tilde K^{(\L)}_{q,{\bf z}}({\cal X})\phi^T({\cal X})\Big|_{{\bf z}=z}
\;,\nn\eea
where $R(r)$ is the multiplicity of $r$ in $R$.
The sums in the first line involve connected rod or polymer configurations containing at least one rod 
centered at $x$; similarly, the sums in the second line involve connected rod or polymer configurations 
containing at least one rod 
centered at $x$ and one rod centered at $y$. All the sums are exponentially convergent and their 
evaluation finally leads to the finite volume analogues of Eqs.(\ref{3.10})-(\ref{3.11}). 
The infinite volume counterparts are obtained simply by replacing all the finite volume activities with 
their infinite volume counterparts and by dropping the constraints that the polymers should be 
contained in $\L$. The infinite volume limit is reached exponentially fast and all the observables 
share the same invariance properties as the infinite volume activities themselves. In particular, 
the infinite volume Gibbs measures $\media{\cdot}^q$ are translation invariant, and the averages
$\media{\c^{-q}_{\x_0}}^q$ and $\media{\prod_{j}n_{x_j}}^q$ are all independent of $q$. We will not 
belabor the proofs of these claims, since they are all straightforward consequences of the cluster 
expansion described in the previous sections, in the same sense as the representations for 
$\media{\c^{-q}_{\x_0}}^q$, $\media{n_x}^q$ and $\media{n_xn_y}^q$ and the proof of their 
convergence, discussed in this section, are a consequence of the bounds of sections \ref{sec4} and 
\ref{sec5}. This concludes the proof of the main theorem.\qed

\vskip1.truecm

{\bf Acknowledgements.} We gratefully acknowledge financial support from the ERC Starting Grant 
CoMBoS-239694. We warmly thank Emanuele Caglioti, for several key ideas and illuminating 
suggestions, which stimulated us to start this project and allowed us to complete it successfully. 
We thank G. Gallavotti, J. Imbrie, J. Lebowitz and E. Lieb for many useful discussions, and
H. Tasaki, for making us aware of this problem.


\end{document}